## Review Article

# Impact of Severe Plastic Deformation on Kinetics and Thermodynamics of Hydrogen Storage in Magnesium and Its Alloys

Kaveh Edalati[1,*], Etsuo Akiba[2], Walter J. Botta[3], Yuri Estrin[4,5], Ricardo Floriano[6], Daniel Fruchart[7,8], Thierry Grosdidier[9,10], Zenji Horita[1,11-13], Jacques Huot[14], Hai-Wen Li[15], Huai-Jun Lin[16], Ádám Révész[17] and Michael J. Zehetbauer[18]

[1] WPI, International Institute for Carbon-Neutral Energy Research (WPI-I2CNER), Kyushu University, Fukuoka, Japan
[2] International Research Center for Hydrogen Energy, Kyushu University, Fukuoka, Japan
[3] Departamento de Engenharia de Materiais, Universidade Federal de São Carlos, Sao Carlos-SP, Brazil
[4] Department of Materials Science and Engineering, Monash University, Clayton, VIC 3800, Australia
[5] Department of Mechanical Engineering, The University of Western Australia, Crawley, WA 6009, Australia
[6] School of Applied Sciences, University of Campinas (UNICAMP), Limeira, São Paulo, Brazil
[7] Institut Néel, CNRS & UGA, 38042 Grenoble, France
[8] JOMI-LEMAN SA, 74890 Fessy, France
[9] Université de Lorraine, Laboratory of Excellence on Design of Alloy Metals for low-mass Structures (DAMAS), Metz, F-57070, France
[10] Université de Lorraine, Laboratoire d'Etude des Microstructures et de Mécanique des Matériaux (LEM3 UMR 7239), Metz, F-57070, France
[11] Graduate School of Engineering, Kyushu Institute of Technology, Kitakyushu, Japan
[12] Magnesium Research Center, Kumamoto University, Kumamoto, Japan
[13] Synchrotron Light Application Center, Saga University, Saga, Japan
[14] Hydrogen Research Institute, Université du Québec à Trois-Rivières, 3351 des Forges, Trois-Rivières, QC G9A 5H7, Canada
[15] Hefei General Machinery Research Institute, Hefei 230031, China
[16] Institute of Advanced Wear & Corrosion Resistance and Functional Materials, Jinan University, Guangzhou 510632, China
[17] Department of Materials Physics, Eötvös University, Budapest, H-1518, P.O.B. 32, Budapest, Hungary
[18] Faculty of Physics, University of Vienna, Boltzmanngasse 5, A-1090 Wien, Austria

*Corresponding author (E-mail: kaveh.edalati@kyudai.jp; Tel/Fax: +81-92-802-6744)

**Abstract**

Magnesium and its alloys are the most investigated materials for solid-state hydrogen storage in the form of metal hydrides, but there are still unresolved problems with the kinetics and thermodynamics of hydrogenation and dehydrogenation of this group of materials. Severe plastic deformation (SPD) methods, such as equal-channel angular pressing (ECAP), high-pressure torsion (HPT), intensive rolling and fast forging, have been widely used to enhance the activation, air resistance, and hydrogenation/dehydrogenation kinetics of Mg-based hydrogen storage materials by introducing ultrafine/nanoscale grains and crystal lattice defects. These severely deformed materials, particularly in the presence of alloying additives or second-phase nanoparticles, can show not only fast hydrogen absorption/desorption kinetics but also good cycling stability. It was shown that some materials that are apparently inert to hydrogen can absorb hydrogen after SPD processing. Moreover, the SPD methods were effectively used for hydrogen binding-energy engineering and synthesizing new magnesium alloys with low thermodynamic stability for reversible low/room-temperature hydrogen storage, such as nanoglasses, high-entropy alloys, and metastable phases including the high-pressure γ-$MgH_2$ polymorph. This article reviews recent advances in the development of Mg-based hydrogen storage materials by SPD processing and discusses their potential in future applications.

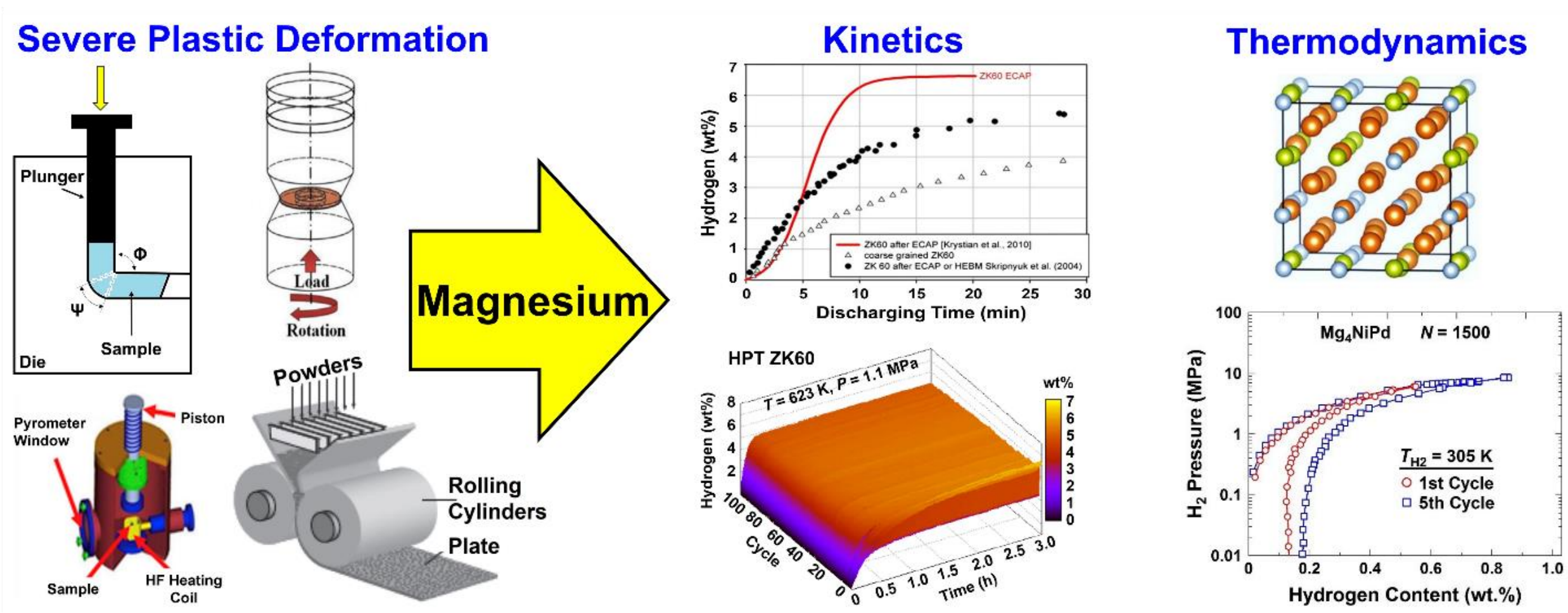

**Table of Contents**

## 1. Introduction

Hydrogen, the lightest element in the periodic table, is considered to be the cleanest fuel of the 21st century. The burning of hydrogen produces only water, making it an appropriate fuel without any negative effects on global warming by $CO_2$ emission. Moreover, the electrochemical reaction of hydrogen with oxygen in fuel cells has higher energy efficiency compared to the combustion of fossil fuels which suggests another advantage of using hydrogen as a fuel [1]. Despite the advantages of hydrogen, there are still drawbacks concerning hydrogen production, utilization and storage that need to be addressed [2].

Hydrogen is mainly produced from gas reforming and a large amount of $CO_2$ is generated during this process. There are significant attempts to produce hydrogen by water splitting using renewable energies via electrolysis and photocatalysis [2], although the efficiency of photocatalysis is still quite low [3]. Fuel cells have high efficiency for the utilization of hydrogen, but reducing their working temperature, developing feasible catalysts, and reducing their price are some major challenges in using fuel cells [4]. Hydrogen can be stored in the form of high-pressure gas, liquid, or solid, but these methods have some limitations [5]. Storage of hydrogen in the form of high-pressure gas needs special tanks due to safety issues and these tanks occupy large space [2,5]. Storage of hydrogen in the liquid form is a compact technology, but liquifying hydrogen consumes energy and liquid hydrogen can evaporate over time [2,5]. Solid-state hydrogen storage, particularly in the form of metal hydrides and complex hydrides, provides the most compact and safest technology for hydrogen storage, but most hydrides suffer either from high thermodynamic stability (i.e. high working temperature), slow kinetics, difficult activation, or low gravimetric storage capacity [5,6].

Magnesium and its hydride $MgH_2$ are the first hydrogen storage materials that were introduced in 1951 [7]. Magnesium is an abundant element in the Earth's crust, it is rather cheap, and it can store 7.6 wt% of hydrogen [8]. However, magnesium exhibits strong Mg-H bonding, which results in great thermodynamic stability of the hydride requiring high hydrogen desorption temperatures well above 573 K [9]. Moreover, it suffers from poor activation and slow hydrogenation/dehydrogenation kinetics mainly due to the presence of an oxide layer on its surface and the slow diffusion of hydrogen in the bulk [8,9]. Alloying of magnesium, development of composites and intermetallic compounds, the addition of catalysts, and microstructural modifications are some strategies used to solve the thermodynamic and kinetic problems of Mg-based hydrogen storage materials [8,9]. Processing by severe plastic deformation (SPD), first proposed by Skripnyuk *et al.* [10], is an effective strategy used within the past two decades to address both the kinetics and the thermodynamics of hydrogen storage in Mg-based materials.

In SPD, a large plastic strain is induced in a piece of material to produce a final product in a bulk form with ultrafine grains or nanograins and large fractions of crystal lattice defects [11]. Equal-channel angular pressing (ECAP) [12], high-pressure torsion (HPT) [13], and accumulative roll-bonding (ARB) [14] are currently the most popular SPD processes, but there are some trends to induce severe strain in hydrogen storage materials by conventional methods such as intensive rolling [15], fast forging [16] and shot peening [17]. It was shown that the presence of a high density of grain boundaries and crystal lattice defects can provide pathways for hydrogen transport to improve hydrogenation kinetics [18]. Unlike the ball milling technique which is traditionally used to produce hydrogen storage materials in the form of powders with a large surface area [8,9], the SPD processes produce bulk samples with a lower surface area. This results in better activation and enhanced air resistance [18]. Furthermore, the SPD processes, particularly those with extremely large shear strains, which are known as ultra-SPD, can be used to synthesize a wide range of Mg-based hydrogen storage materials even from immiscible systems [19].

In the present article, recent advances in the application of SPD to Mg-based hydrogen storage materials are reviewed with a focus on processing, kinetic features, and thermodynamic modification via the synthesis of novel materials.

## 2. Processing Methods of Mg-Based Materials

Although numerous SPD methods have been developed to process structural materials with advanced mechanical properties [11,12], only a limited range of these methods have been used for processing hydrogen storage materials. The main impact of these methods is the refinement of microstructure and the introduction of crystal lattice defects such as vacancies and dislocations. Among the various SPD methods, HPT is capable of continuously inducing large strains under high pressure and is the most efficient one for microstructure refinement and defect generation in Mg-based compounds [13]. Since repetitive cycles and discontinuous straining are applied in ECAP and intensive rolling, the processing strain and the concomitant microstructure refinement are not as significant as HPT. The advantage of the former techniques is their suitability for processing large pieces of hydrogen storage materials, which is beneficial for commercial applications [11]. Fast forging can also induce a large strain within one cycle, but the amount of strain and the degree of grain refinement by this method are lower compared with those enabled by HPT processing [16]. In this section, the major studies conducted on SPD processing of Mg-based hydrogen storage materials using ECAP, HPT, intensive rolling and fast forging are reviewed.

### 2.1. Equal-Channel Angular Pressing

In relation to the efficacy of the processing of magnesium alloys aiming at improving their hydrogen storage properties, ECAP takes a special place among the techniques of SPD. Not only is ECAP arguably the most popular SPD method [20-25], but it is also historically the first one whose potency as a tool for accelerating the hydrogenation kinetics of magnesium alloys by SPD was discovered [10]. In the ECAP method, a billet in the form of a rod or bar is repeatedly pressed through a channel with a bending angle to introduce simple shear strain, as shown in Fig. 1a [20-22]. The ability of ECAP to produce extreme grain refinement, down to submicron scale, in magnesium and its alloys is well documented [22,26-29]. The ECAP-induced acceleration of hydrogen absorption/desorption was first demonstrated for Mg-based alloy ZK60, Mg-4.95Zn-0.71Zr (wt%), which is widely used for structural applications [10]. The effect of ECAP on the dehydrogenation kinetics is shown in Fig. 1b. The data obtained for ZK60 by employing a different ECAP facility, which represent a further improvement over the original results [10], are also shown in Fig. 1b [30]. A striking drop in the particle size down to the submicron range for the material processed by ECAP and subjected to hydrogenation and dehydrogenation is shown in Fig. 1c. Variants of the ZK60 alloy were the object of studies where a combination of ECAP with cold rolling [31] or ARB [32] was used, albeit with lesser success.

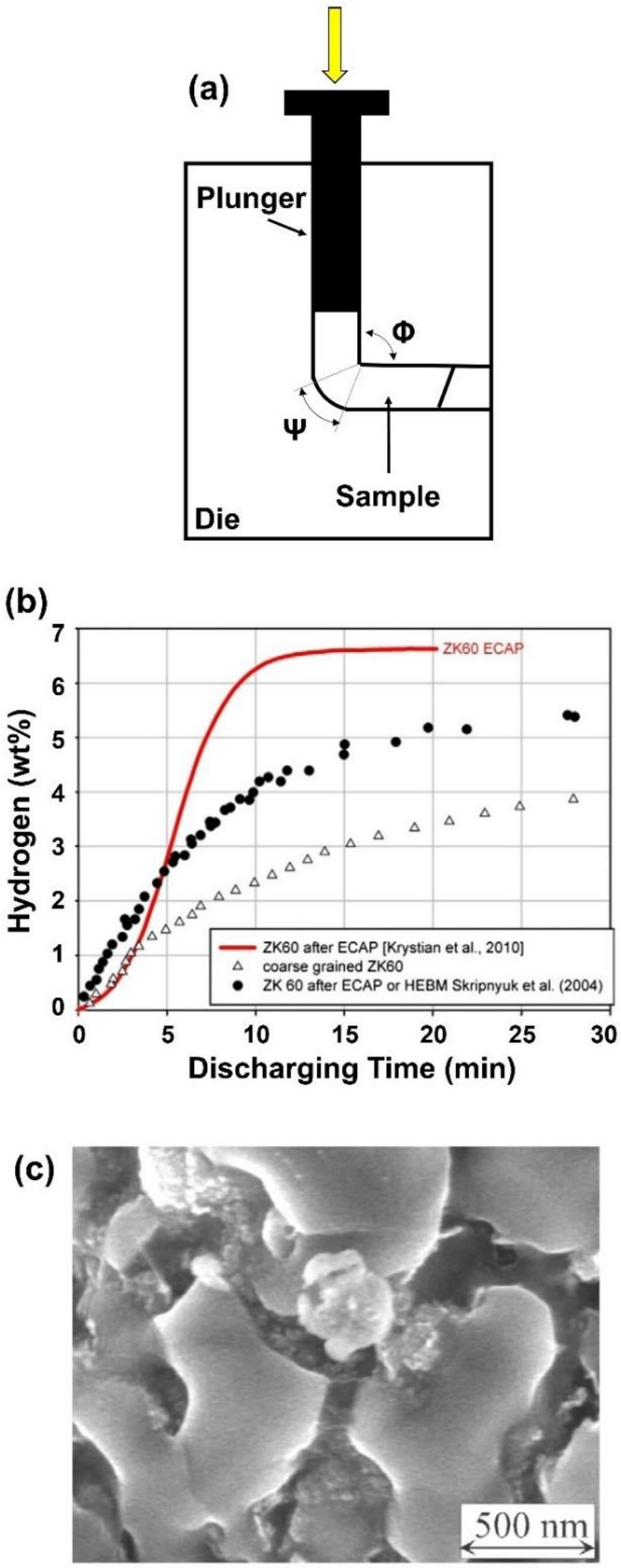

Figure 1. (a) Schematic illustration of ECAP. (b) Dehydrogenation kinetics of alloy ZK60 after different kinds of processing by ECAP (triangles) [10], ECAP followed by high-energy ball milling (closed circles) [10] and ECAP with a different route (red curve) [30]. (c) Scanning electron microscopy images of alloy ZK60 processed by ECAP after hydrogen absorption/desorption cycle [10].

Research conducted so far demonstrated that ECAP is on par or can even outperform the more conventional ball milling technique, both methods improving the hydrogenation kinetics. Indeed, the hydrogenation curve for the ball-milled ZK60 alloy was shown to be very close to, yet lying beneath, the curve measured for the ECAP-processed one [10]. The great benefit of ECAP is the possibility to obtain a promising hydrogen storage material in bulk, with an additional advantage of avoiding potentially hazardous ball milling of powders. Several studies utilizing ECAP as the processing route [33-36] and further SPD techniques were also applied to a range of Mg-based alloys [37,38], notably the eutectic $Mg_{89}Ni_{11}$ alloy with a fine lamellar structure [39].

A promising approach to enhancing the hydrogenation/dehydrogenation kinetics of magnesium is the use of composites containing metal hydrides [40] or particles of carbonaceous materials [41]. Specifically, a study of a Mg-based composite with 2 wt% of multiwall carbon nanotubes processed by ECAP showed that the addition of nanotubes to magnesium leads to a substantial acceleration of hydrogen desorption rate, a further advantage being the disappearance of pressure hysteresis [42]. Viewed as a progenitor of SPD-based tools for improving the hydrogenation kinetics of Mg-based alloys, ECAP processing can be regarded as a viable avenue to large-scale hydrogen storage facilities [43].

### 2.2. High-Pressure Torsion

Among different bulk SPD methods, HPT [13] has successfully been applied to manufacture a large variety of different hydrogen storage materials [43-47] due to the exceptionally high shear strain that can be achieved in bulk sample volume [21,48]. During HPT deformation, as schematically shown in Fig. 2a, a disc-shaped specimen is inserted between two anvils and subjected to concurrent uniaxial pressure of several Gigapascals and torsional straining by several revolutions [49]. There has been a wide range of applications of HPT to Mg-based materials, as will be discussed here.

HPT deformation of $MgH_2$ powders results in a significant grain refinement [50] and also induces a strong (002) texture and the formation of the metastable $\gamma$-$MgH_2$ phase [51] which positively influences the overall hydrogen storage performance. The particle size of magnesium powders has a strong effect on the hydrogen absorption kinetics, but the grain/crystal size is also known to affect the kinetics significantly [52]. It was reported that a bimodal microstructure develops when bulk magnesium is subjected to HPT, including the emergence of nanocrystals and large recrystallized grains, resulting in a substantial improvement of the hydrogenation kinetics [53].

The average density of dislocations in commercial magnesium processed by HPT reaches a very large value of $8 \times 10^{15}$ $m^{-2}$, which can act as hydrogen transport sites [54]. A ZK60 Mg-based alloy processed by HPT exhibits a stable storage capacity of up to 100 hydrogenation cycles, as shown in Fig. 2b and 2c [55]. The maximum hydrogen capacity of ball-milled Mg-Ni nanopowder increases by 50% after HPT and reaches the theoretical value, due to the creation of new hydrogen transport sites in the vicinity of dislocations [56], while the formation of $Mg_2NiH_{0.3}$ hexagonal solid solution and the monoclinic $Mg_2NiH_4$ takes place [57-58]. Other lattice defects, notably stacking faults, can also improve the hydrogenation kinetics of $Mg_2Ni$ processed by HPT [59]. A large fraction of cracks in ultrafine Mg + 2 wt% Ni powder can act as pathways for hydrogen transport from the surface of the HPT-processed disc [60]. As will be discussed later, the extreme shear deformation during torsion can reach such a high magnitude that it is capable to promote hydrogen uptake even in the non-absorbing $MgNi_2$ phase [61].

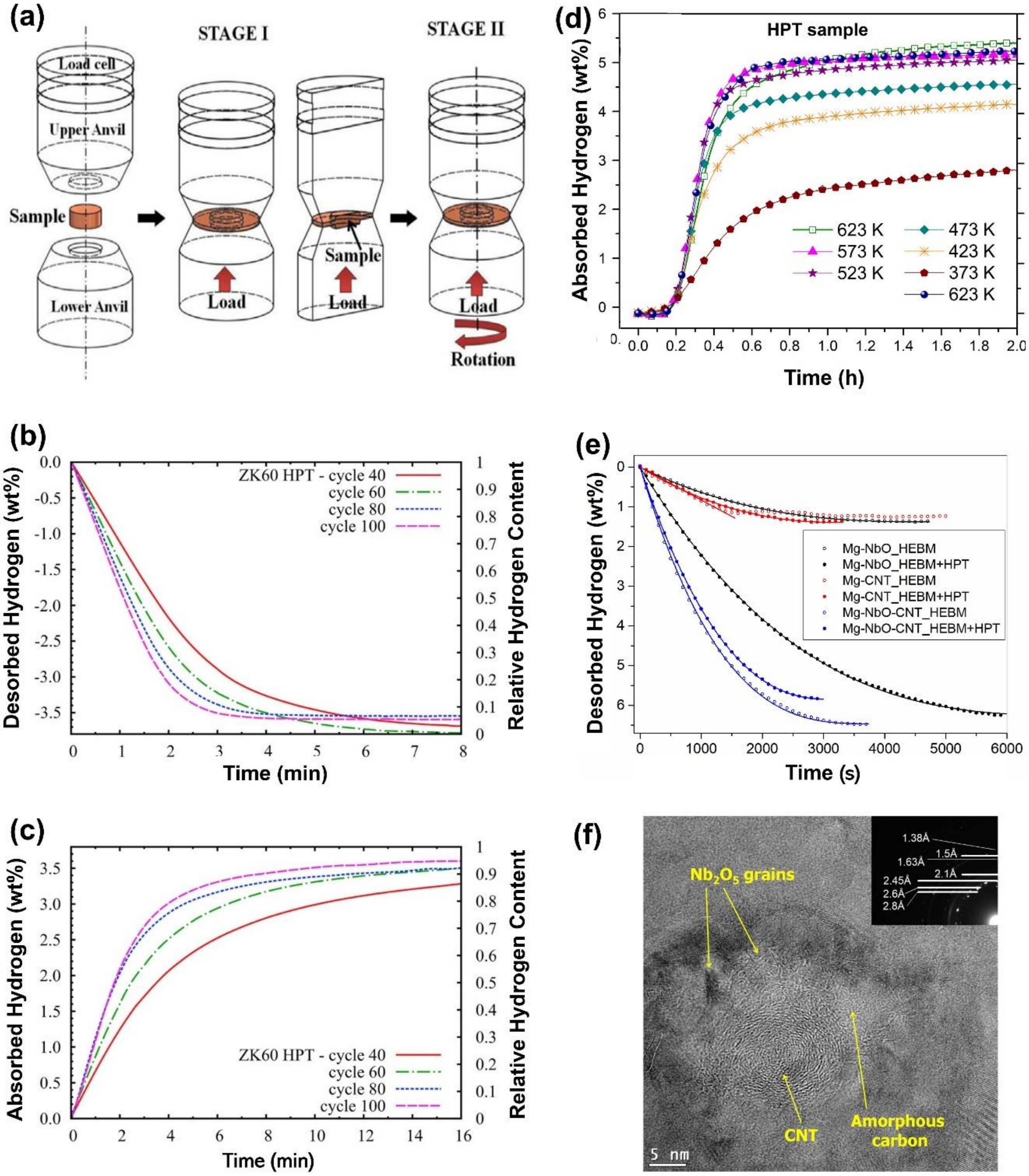

Figure 2. (a) Schematic representation of HPT apparatus with uniaxial compression and simultaneous torsion [49]. (b) Dehydrogenation and (c) hydrogenation kinetic curves obtained at 623 K and 1 MPa absorption pressure, and at 10 Pa desorption pressure for HPT-deformed ZK60 alloy [55]. (d) Hydrogenation curves at different temperatures for Mg + 5 wt% Ni + 2 wt.% $Nb_2O_5$ powder composite processed by HPT [62]. (e) Desorption kinetic curves obtained at 573 K and 1 kPa for as-milled magnesium powders catalyzed by $Nb_2O_5$ and/or carbon nanotubes and corresponding HPT-processed discs. (f) High-resolution lattice image of cycled HPT-processed Mg + $Nb_2O_5$ + carbon nanotube (inset: selected area electron diffraction pattern) [65].

The HPT process can be used effectively to fabricate new Mg-based composites or alloys for hydrogen storage. A reasonable hydrogenation capacity can be obtained for a powder mixture of Mg + 5 wt% Ni + 2 wt% $Nb_2O_5$ subjected to HPT at a temperature as low as 423 K, as shown in Fig. 2d [62]. It was revealed recently that metastable phases can be developed even in immiscible systems (such as Mg-V-Sn [19], Mg-V-Ni [19], Mg-Ti [63] and Mg-Zr [64]) during ultra-SPD, thus new potential hydrogen storing materials can be manufactured. It was found in a recent study that the combination of ball milling and the HPT process can further improve the desorption kinetics of nanocrystalline magnesium catalyzed by $Nb_2O_5$ and/or carbon nanotubes [65,66]. As confirmed by transmission electron microscopy, the carbon nanotubes acting as diffusion channels for hydrogen are preserved during plastic deformation by ball milling and HPT as well as during sorption cycling, as shown in Fig. 2e and 2f [65]. The combined catalytic effect of metal-oxide particles and carbon nanotubes can be substituted by applying only metal-oxide nanotubes [67].

HPT can also be utilized for the synthesis of high-entropy materials, such as MgVTiCrFe for hydrogen storage [68]. Moreover, the hydrogenation performance of fully disordered systems, like glassy alloys, can be improved significantly when the materials are subjected to severe shear deformation, including a reduced hydrogenation temperature and improved hydrogen sorption kinetics for $Mg_{65}Ni_{20}Cu_5Ce_{10}$ [69] and $Mg_{65}Ni_{20}Cu_5Y_{10}$ [70] metallic glasses. At the same time, the hydride formation enthalpy increases noticeably which is especially more pronounced in the most deformed perimeter region of the HPT-processed discs [71]. Despite the potential of HPT to be applied to a wide range of materials, its main limitation is currently the small size of the sample which is an obstacle to its development for commercial applications.

### 2.3. Intensive Rolling

Despite the successful use of high-energy ball milling methods to produce nanostructured Mg-based alloys, two processing techniques of ECAP schematically shown in Fig. 1a [10] and intensive rolling schematically shown in Fig. 3a-c [72], were reported almost at the same time, as alternatives enabling the production of "bulk" samples with refined and more air-resistant microstructures. Metal processing based on cold rolling, such as co-lamination via repetitive rolling, was first used in hydride-forming Mg-based alloys to produce Mg-Ni composite structures [72]. Heat treatment of a deformed sandwich of Mg and Ni foils resulted in the formation of an intermetallic $Mg_2Ni$ compound. A similar result was observed for Mg-Al composite alloys with the final heat treatment forming the $Mg_{17}Al_{12}$ compound that also absorbed hydrogen reversibly [73]. Cold rolling of a stack of Mg/Pd foils produced an air-resistant laminated compound with a shorter activation time compared to a ball-milled sample of the same composition as well as to pure magnesium, as shown in Fig. 3d [74]. Repetitive rolling was also used to prepare laminate composites of Mg/Cu [75,76] and Mg/Pd [75] with the micrometer-order layered structure containing a high density of dislocations and vacancies. A similar route to co-lamination, but involving a bonding process during intensive rolling, designated as ARB shown schematically in Fig. 3a [77], was used to prepare Mg/Ti multilayers and Mg/stainless steel composites [78]. Activation for hydrogen absorption improved with an increased number of fold and roll operations, which caused increasing refinement of the hard second-phase particles.

Intensive cold rolling was effectively used to refine the microstructure and disperse the particles of catalysts. Intensive rolling was used to process mixtures of $MgH_2$-Fe as an alternative to reach the grain size reduction typically obtained by ball milling [79,80]. Cold rolling of commercial $MgH_2$ was considered equivalent to ball milling in terms of microstructure refinement [81], as also observed for Mg-$LaNi_5$ composites [82]. Figs. 3b and 3c show the schematics of a vertical and horizontal rolling machine used to process $MgH_2$ powders, respectively [34]. The presence of additives in Mg-based alloys processed by cold rolling proved to be as efficient as ball-milled materials despite the distribution of the additive particles being much less homogeneous. A uniform distribution of additive particles was also observed in $MgH_2$ powders containing different types of additives [83]. It was shown that the uniformity of $FeF_3$ distribution in $MgH_2$ powders can be improved by a combination of short-time ball milling followed by intensive rolling [84].

There have been some improvements in the rolling process for treating hydrogen storage materials. The use of a protective atmosphere during rolling allows further refinement of the microstructure in commercial magnesium [85] and $MgH_2$ [86] since a greater number of rolling passes can be applied without surface contamination. Fig. 3e shows the refined and textured microstructure achieved in $MgH_2$ after 35 rolling passes under an argon atmosphere which can result in fast hydrogenation and dehydrogenation kinetics. Figs. 3f and 3g show the positive effect of the number of rolling passes under a protective atmosphere on the kinetics of hydrogen absorption and desorption in $MgH_2$ [86]. The same route was used to process $MgH_2$-$LaNi_5$ mixtures under a protective atmosphere, resulting in compacted composite flakes [87], which exhibited faster hydrogen absorption kinetics and reduced desorption temperatures in comparison with single-phase $MgH_2$. Low-temperature rolling in AZ91 alloy can improve the hydrogen storage properties compared with rolling at ambient temperature due to the larger density of microcracks and consequently high density of exposed interfaces and formation of stronger (002) texture [88].

Cold rolling has also been used as a further processing step to improve the activation and hydrogen absorption kinetics of Mg-based samples initially prepared by a different technique. It was shown that cold rolling following other modes of SPD processing of magnesium and its alloys results in a rolling texture that can enhance the kinetics in the first hydrogenation cycle by exposing less densely packed atomic planes to the hydrogen atmosphere [24]. Studies of other processing combinations, such as cold rolling of melt-spun ribbons [89], cold rolling of HPT-processed discs [90], and ARB processing of ECAP-treated ZK60 alloy [32] demonstrated that in addition to the refined microstructure, the presence of free surfaces (or interfaces) and texture have positive effects on the hydrogen storage properties of Mg-based alloys. Intensive rolling is likely to have the greatest potential for large-scale processing of hydrogen storage materials, although the magnitude of plastic deformation in this method is smaller than in ball milling followed by HPT.

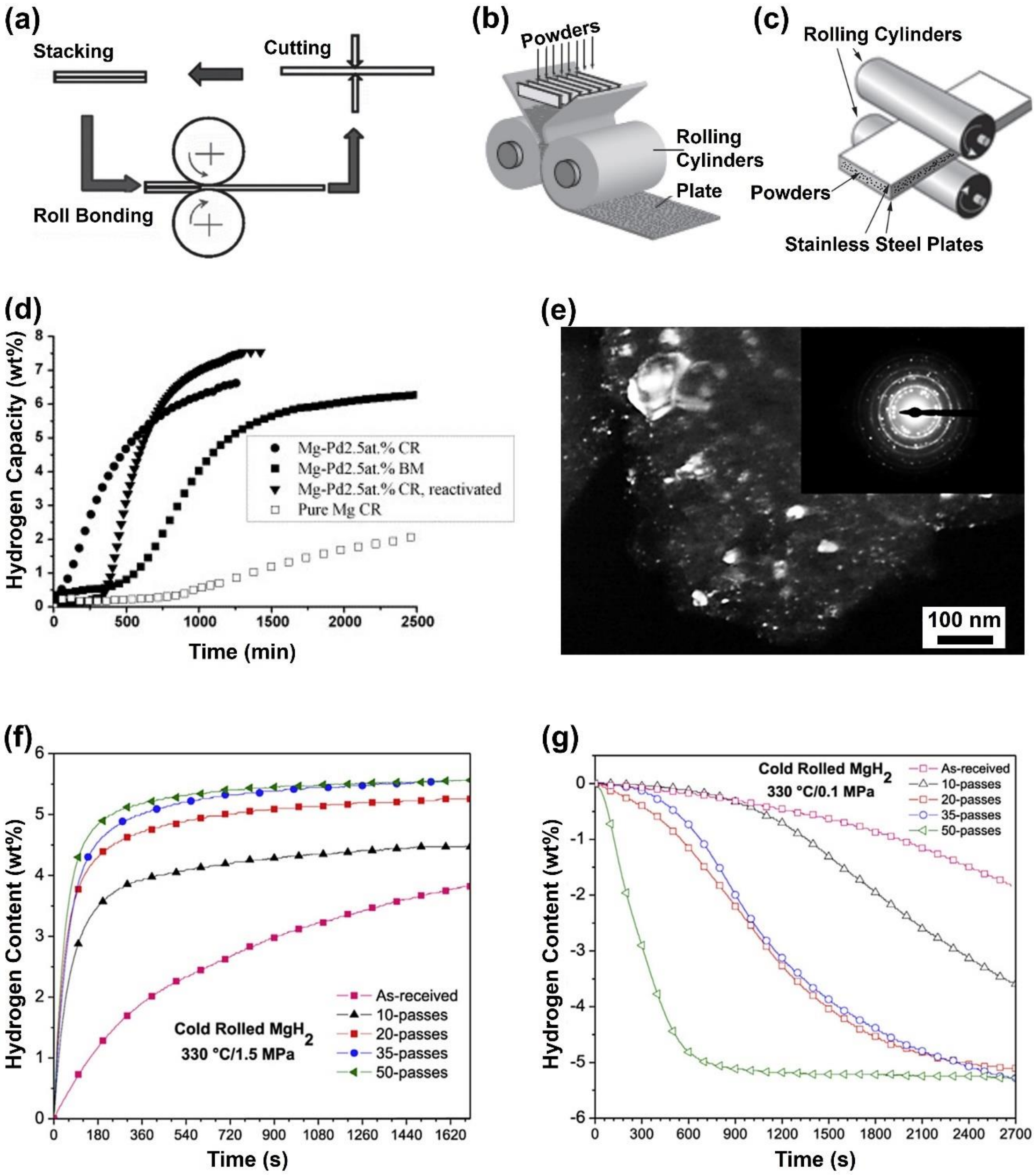

Figure 3. (a) Schematic illustration of ARB process [77], and sketches of (b) vertical and (c) horizontal rolling machines used to process $MgH_2$ powders [34]. Hydrogen absorption curves under a hydrogen pressure of 1.5 MPa at 623 K for pure magnesium and Mg - 2.5 at% Pd after cold rolling (CR) and ball milling (BM) [74]. (e) Dark-field image and corresponding selected area electron diffraction pattern, taken by transmission electron microscopy, for $MgH_2$ processed by cold rolling with 35 cycles. (f) Hydrogen absorption and (g) desorption curves for $MgH_2$ processed by different numbers of passes of cold rolling under a protective atmosphere [86].

### 2.4. Fast Forging

As discussed earlier, the kinetic performance of hydrogen storage Mg-based materials can be improved by: (i) reducing the particle/crystallite sizes to the nanometer level combined with including a high lattice strain and a large density of extended defects (dislocations, stacking faults, twins, etc.) by plastic deformation (high-energy ball milling [91-96], HPT [24,35,44,50,51,97,98], ECAP [10,33,99,100-102], intensive rolling [81,103,104]), and (ii) using various additives which can act as catalysts to accelerate hydrogen sorption [104-108]. In any case, mass production of $MgH_2$ must satisfy practical requirements such as processing conditions (ease of manufacture, efficiency, cost, etc.) and material performance (maximum hydrogen uptake, fast sorption, stability, lifetime, etc.) [109]. These requirements can be achieved using fast forging, a less conventional SPD method [16,110,111].

A photograph of the fast-forging facility is shown in Fig. 4a. In this method, the height of ingots of bulk magnesium, Mg-based alloys or their compacted composites is reduced by about a factor of ten within about $5 \times 10^{-3}$ s, while the initial temperature of the sample is set using an induction high-frequency coil. Since the mechanical energy of the hammer is almost entirely dissipated in the material without elastic rebound, it leads to refining the grain size plastically, while simultaneously healing the sample frictionally [16,110,111]. Such grain refinement by fast forging can improve the hydrogenation kinetics similarly to other SPD methods.

In addition to microstructural modification, fast forging can be used to synthesize composites of Mg-based hydrogen storage materials. As a trial, homogeneous mixtures of strongly compacted powders of Mg+Ni with initial particle sizes of 5-30 µm for magnesium and 30-40 µm for nickel were fast forged. To optimize forging conditions and determine the deformation fields, a two-dimensional calculation model was developed by considering the particle size, distribution of nickel among magnesium particles and their relative hardness [112]. Numerical simulation of the adiabatic compression processes quantifies a marked increase of sample temperature for average strains of 80-90%, while the additional heat generated by an increase in strain was found to drop [113,114]. Fig. 4b shows the conversion rate of $Mg_{89}/Ni_{11}$ powder mixtures to form $Mg_2Ni$. The examination of absorption traces at 614 K and 2 MPa of $H_2$ indicate that fast forging of magnesium in a brittle state at low temperature promotes the formation of defects and cracks enhancing hydrogen diffusion in the bulk, while forging at high temperature and a ductile state of the material allows the formation of $Mg_2Ni$ as a catalyst, as reported for various SPD routes [39,114-118].

In another study, compacts of 95 wt% Mg + 5 wt% $MgH_2$ were processed by combining fast forging with two passes of ECAP at room temperature. Pressure-composition-temperature isotherm analyses showed that at 593 K and under a hydrogen pressure of 1 MPa, the fast forging was effective to achieve up to 7 wt% of hydrogen uptake in ~140 min, as shown in Fig. 4c [119]. The performance characteristics were found to be rather similar to those of the ECAP-processed samples with 6 wt% hydrogen uptake at 593 K and 1 MPa in 60 min [120,121]. Such improvements by fast forging are also comparable with those achieved using other SPD routes [122,123]. These results confirm that the fast-forging process can be considered as an efficient SPD route enabling mass production of $MgH_2$ for hydrogen storage [124].

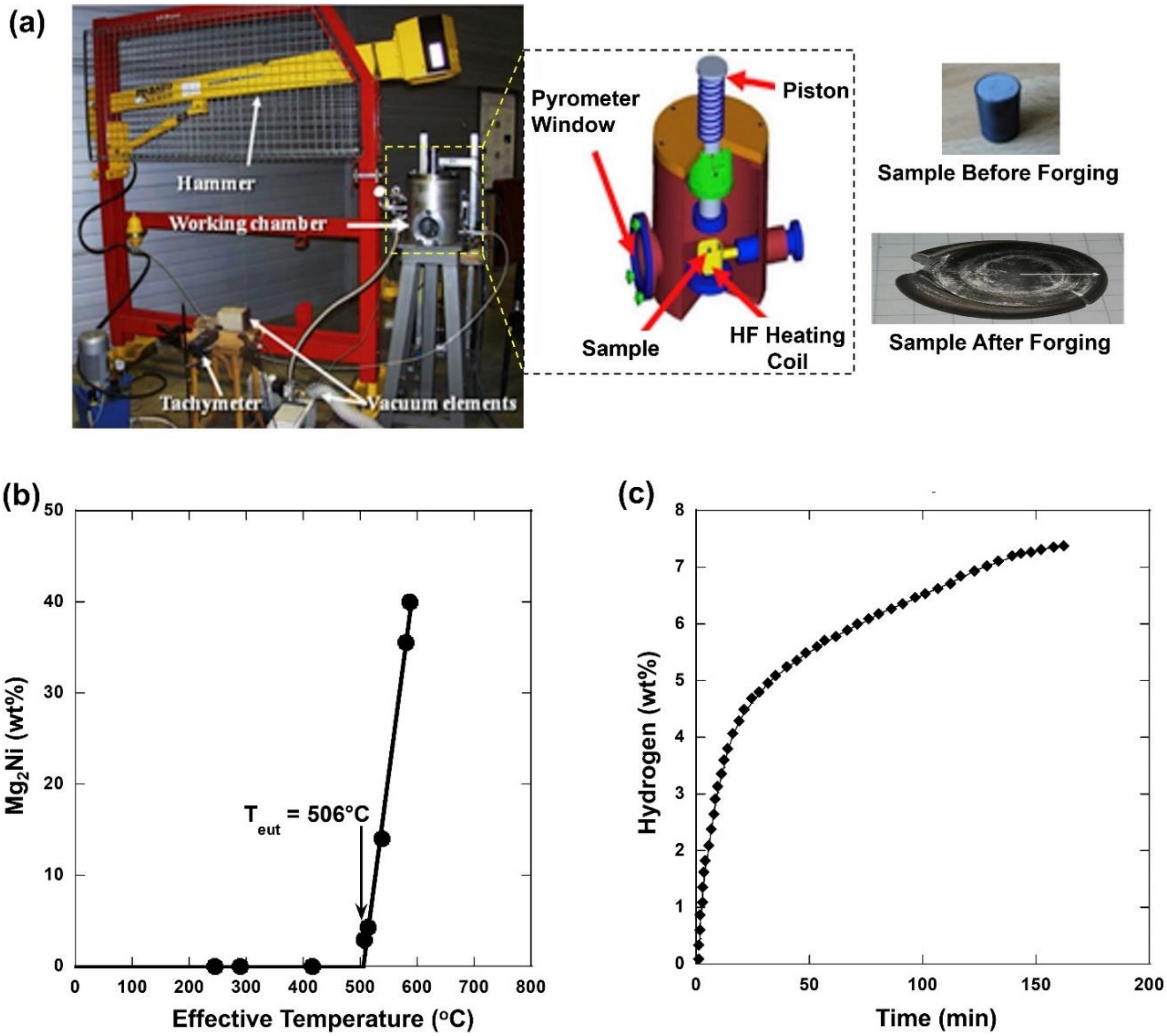

Figure 4. (a) Fast forging facility with controllable atmosphere and temperature having a 150 kg hammer falling from up to 1.5 meters with a forging time of less than 0.02 s (left), schematic illustration of working chamber (center) and the appearance of samples before and after forging (right) [16]. (b) The conversion rate of $Mg_{89}/Ni_{11}$ powder mixture to $Mg_2Ni$ after fast forging at different temperatures [114]. (c) First hydrogen absorption kinetic curves at 593 K and under a hydrogen pressure of 1 MPa for 95 wt% Mg + 5% wt% $MgH_2$ compact processed by fast forging [119].

## 3. Kinetic Features of Severely Deformed Mg-based Materials

As mentioned above, many investigations have been done on the hydrogenation/dehydrogenation kinetics of metal hydrides. The popular ways to enhance kinetics are the addition of a catalyst or change of stoichiometry [125-127]. Other efficient ways to enhance kinetics discussed in several sections of the present review are mechanical deformation of powders [128] or bulk samples [129-132]. However, for most practical applications, dehydrogenation can be relatively slow, and thermodynamics (plateau pressure) is more crucial than kinetics. For the hydrogenation part, because of the high enthalpy of the formation of the hydrides, the main rate-limiting step will be the heat transfer. For example, storing one kilogram of hydrogen in magnesium hydride for six minutes will require a heat transfer of about 100 kW. Therefore, for practical hydrogen tanks made of metal hydrides, the main challenge will be to manage a rapid heat transfer. In addition, for commercial applications, the cost of the metal hydride should be as low as possible.

Despite the significance of plateau pressure, heat transfer and cost, it is still essential to find strategies to address the kinetic drawbacks of hydrogen storage materials. SPD processing generates a high density of crystal lattice defects and refines the microstructure of magnesium and its alloys and as a result improves its hydrogenation kinetics, while keeping its reversibility. In contrast with other methods used to enhance hydrogenation kinetics, such as alloying, catalyst addition, synthesis of composites, and amorphization [133,134], the SPD methods do not need any extra additives. Furthermore, as opposed to ball milling which produces potentially hazardous fine powders [133,134], the SPD-processed materials are in a bulk form with less contact with the air atmosphere. In addition, they have a large density of crystal lattice defects, which can act as fast hydrogen pathways for easy activation. These kinetic features of severely deformed materials can sometimes result in hydrogen uptake in materials that are normally inert to hydrogen. These kinetic features are briefly reviewed in this section.

### 3.1. Activation

One issue that is often overlooked is the problem of the activation of hydrogen storage materials. For metal hydrides, the term activation means the process by which the alloy is prepared for reversible hydrogen sorption [135]. Typically, it involves submitting the alloy to hydrogen pressure and/or temperature much higher than the ones predicated by thermodynamics. However, this implies an additional cost for the manufacturer because the storage tank needs to be overdesigned so that it can sustain not only the service conditions but also the activation conditions. The ideal situation would be to perform the activation process under the same regime as the service conditions of the storage tank. The problem of activation has been investigated for a long time, but it is still not clear what mechanisms are at play [136]. What is known is that microstructure plays a crucial role in the activation of metal hydrides. Mechanical deformation has been shown to improve the activation kinetics of metal hydrides [99]. The most widely used method is ball milling [137,138], but recently other mechanical methods or a combination of them have been shown to improve the activation of metal hydrides. Some of them are cold rolling [74,89,139,140], ECAP [30,31], and HPT [59]. As examples of the effectiveness of mechanical deformation, Fig. 5 shows the effect of HPT on (a) microstructure and (b) the first hydrogenation of magnesium [53], and (c) the effect of ECAP passes on the first hydrogenation of AZ61 magnesium alloy [141]. Clearly, mechanical deformation introduces ultrafine grains and crystal lattice defects and accelerates the first hydrogenation.

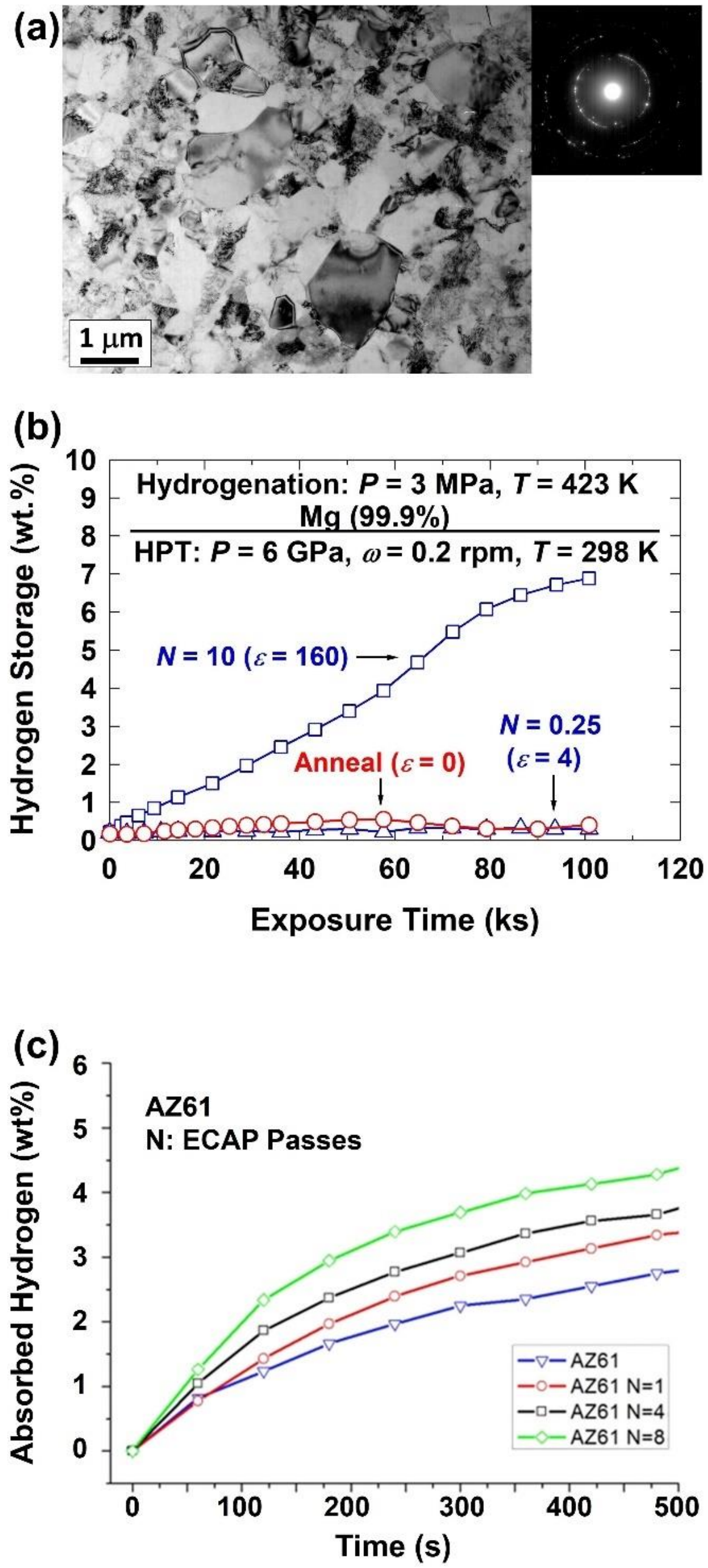

Figure 5. (a) Microstructure and corresponding selected area electron diffraction of magnesium processed by ten turns of HPT ($N$ = 10) [53]. (b) First hydrogenation (activation) of magnesium at 423 K and under 3 MPa of hydrogen before and after processing by HPT for $N$ = 0.25 and $N$ = 10 turns [53]. (c) First hydrogenation of AZ61 alloy at 643 K and a hydrogen pressure of 3.5 MPa before and after processing by ECAP for $N$ = 1, 4 and 8 passes [141].

### 3.2. Air Resistance and Hydrogenation Kinetics

The kinetics of hydrogen absorption and desorption in magnesium is slow, and thus, this issue should be addressed by increasing the surface area, generation of grain boundaries, generation of interphase boundaries, alloying with transition metals (Tm = Co, Fe, Mn, Ni, Nb, Ti, V), and the addition of catalysts. The milling process in the form of high-energy ball milling, mechanical alloying and reactive ball milling (i.e. milling under a hydrogen atmosphere) is effective to induce all these effects [40,92,126,142,143]. However, in some cases, the milling process does not necessarily improve the kinetics because the generated active surfaces can react easily to be oxidized during processing or subsequent handling. In addition, because of this high surface reactivity, the produced powders are pyrophoric which can create significant safety issues. Thus, the ball-milled magnesium powders should be handled under a controlled atmosphere such as in the glove box in the laboratory. Therefore, despite the efficiency of ball milling and particularly reactive ball milling in mechanical alloying, uniform addition of catalysts, grain size reduction, and generation of high densities of grain boundaries and defects [50,144,145], the difficulties in handling the active powders under air atmosphere have greatly restricted its potential applications. Several SPD techniques can produce similar benefits as ball milling in enhancing the hydrogen sorption kinetics with an additional advantage of the bulk materials processed by these techniques being more air resistant and less contaminated compared with ball-milled powders [98]. The low surface area of bulk materials processed by SPD compared to powders allows for storing them under an air atmosphere for a long time. The presence of crystal lattice defects, particularly grain boundaries, provides numerous pathways for hydrogen transport from the partially oxidized surface to the bulk to produce the hydride phase [18]. Moreover, it was shown that the consolidation of powders to a compact material using SPD methods not only enhances the air resistance but also accelerates the hydrogen storage kinetics [52,60,146]. These SPD methods for additional powder consolidation include intensive extrusion [139,147], intensive rolling [58,139], HPT [44,63,148,149], ECAP [36,150], accumulative fold-forging [151] and fast forging [118].

HPT is often selected for powder consolidation of hydrogen storage materials because it induces large shear strains combined with high pressure. A comparative study of HPT imparted on (i) micro-sized atomized magnesium powder and (ii) nano-sized condensed magnesium powder showed an additional impact of SPD on the kinetics of absorption/desorption [52]. The microstructures of these samples before and after HPT processing are shown in Figs. 6a-d, illustrating the size of the powder precursors and the SPD-induced consolidation. The examination of the first hydrogenation kinetics at 673 K under a hydrogen pressure of 3.5 MPa, shown in Fig. 6e, indicates that the finer condensed magnesium powder has faster hydrogenation kinetics than the atomized one [60,146]. For both types of powder precursors, HPT processing has the effect of improving the hydrogenation kinetics through the introduction of structural defects, microstructure refinement, and breaking of the powder surface oxide layer that was redistributed inside the microstructure to act as a possible catalyst [52]. Another advantage of HPT consolidation is the reduction of the hysteresis between the absorption and desorption plateau pressure during the pressure-composition isotherm experiments [52]. It should be noted that the apparent lower hydrogen uptake capacities after HPT processing in Fig. 6e are due to the large scale of the diffusion path in the bulk HPT samples being rapidly surrounded by compact $MgH_2$, which does not allow the samples to react completely with hydrogen. This effect is reduced in subsequent cycles due to hydrogen-induced fragmentation of samples and diminished diffusion path.

In summary, in addition to the enhancement of the kinetics of hydrogen storage, the consolidation of powders by SPD can also have further benefits. First, the initial powder can be mixed with catalysts almost without any limitations. Second, the produced bulk materials are air-resistant and can be stored under atmospheric conditions [152-154]. Third, the absorption and desorption processes occur usually much faster than in non-consolidated precursor powders [52,139]. Fourth, the microstructural features such as the crystallite size, crystal lattice defects and texture, which were suggested to influence the hydrogenation kinetics [148,150,153], can be easily tailored by controlling the imposed strain and the heat treatment [118].

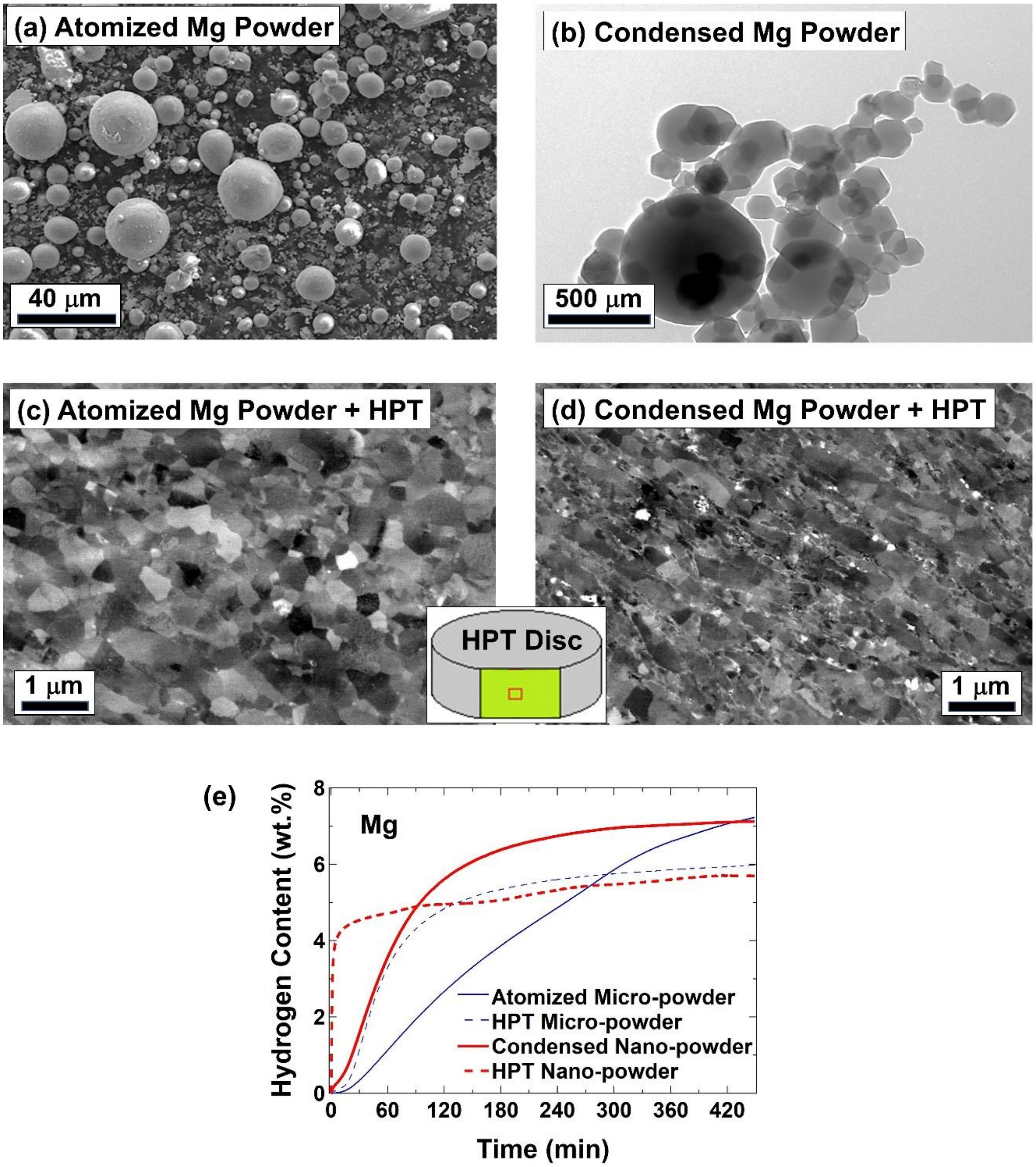

Figure 6. Microstructure of (a) atomized magnesium micro-sized powder, (b) condensed nano-sized powder, (c) HPT-consolidated atomized powder and (d) HPT-consolidated condensed powder, including (e) kinetics for their first hydrogenation [52].

### 3.3. Cycling Stability

Over the last two decades, numerous investigations of hydrogen storage in mechanochemically treated magnesium and Mg-based alloys, including SPD-processed ones, have been reported [130]. The top-down SPD methods have been successfully applied to increase absorption/desorption kinetics with the additional advantage of replacing highly expensive and contamination/oxidation-risky bottom-up methods such as ball milling [10]. Moreover, a combination of SPD methods and *ab initio* calculations have been applied to decrease the desorption temperature by substitutional mechanical alloying of magnesium with elements of optimized type and concentration [129]. The kinetics of hydrogen storage is very sensitive to numerous parameters such as the surface and morphology of the storage material (i.e. oxide layers, porosity, texture, microstructure, and the density/type of SPD-induced crystal lattice defects), and these parameters may change in different experimental procedures or during absorption/desorption cycles [130]. An examination of publications on SPD processing of Mg-based hydrogen storage materials indicates that the results of investigations do not always match unless all the parameters mentioned above are carefully controlled to have a reliable understanding of the hydrogen dissociation (see, e.g. [155]), diffusion and storage processes. Another issue with these publications is that most of them only studied one or a few absorption/desorption cycles, while the stability of the hydride during high-cycle hydrogen storage is a critical issue.

The hydride formation stability is intimately connected with the thermal stability of the microstructure with regard to the temperature necessary for a minimum time to complete absorption/desorption. While in systems with high melting temperatures, the microstructure is stable because of a much lower absorption/desorption temperature compared to the melting point (e.g. TiFe [130]), this is not necessarily true for materials with medium and low melting temperatures, like magnesium. Here, stability can be reached by creating alloys or by adding nanoparticles of a second phase that can stabilize the microstructure beyond the hydrogenation temperature and act as nucleation sites for hydride [156]. Krystian *et al.* [30] for the first time reported this stability for up to 1,000 cycles for ECAP-processed Mg-based ZK60 alloy, but not for ECAP-processed pure magnesium. Nevertheless, Chiu *et al.* [157] confirmed the stability of another ECAP-processed Mg-based alloy, AZ31. Grill *et al.* [55] not only reported highly stable storage characteristics for HPT-processed ZK60 alloy but also found that SPD-induced second-phase particles can act as nuclei for heterogeneous hydride formation, according to a thorough Johnson-Mehl-Avrami analysis [158] yielding an Avrami exponent of $n = 1$. Popilevsky *et al.* [159] reported something similar in the case of composites of magnesium with multiwall carbon nanotubes where they applied some mechanochemical treatment and achieved chains of carbon nanoparticles that became nucleation sites for the hydrogenation. For SPD-processed Mg-based alloys, through combined differential scanning calorimetry and X-ray diffraction analyses, Ojdanic *et al.* [160], Cengeri *et al.* [161], and very recently Abbasi *et al.* [162] found thermally stable SPD-induced vacancy agglomerates [160-162] which heterogeneously nucleate the hydride phase, instead of second-phase precipitates.

Fig. 7a shows the hydrogen absorption rate in a Mg-0.5Zn-1Gd-0.62Y-0.67Nd (wt%) alloy processed by HPT and ECAP after ten hydrogenation cycles. The hydrogenation kinetics in both samples remain fast even after ten hydrogenation cycles, although the ECAP-processed sample shows slightly better performance. Fig. 7 demonstrates the long-cycle hydrogenation (up to 100 cycles) in ZK60 processed by (a) HPT and (b) friction stir processing (FSP), confirming that both samples exhibit high cycling stability. From all these long-cycle storage characteristics shown in Fig. 7, FSP gave rise to the best hydrogen storage kinetics and the best storage capacity, followed by those of ECAP and HPT. Future investigations should be conducted to understand how the type and/or density of the vacancy agglomerates depend on the applied SPD method and thus govern the long-cycle hydrogenation in different Mg-based compounds.

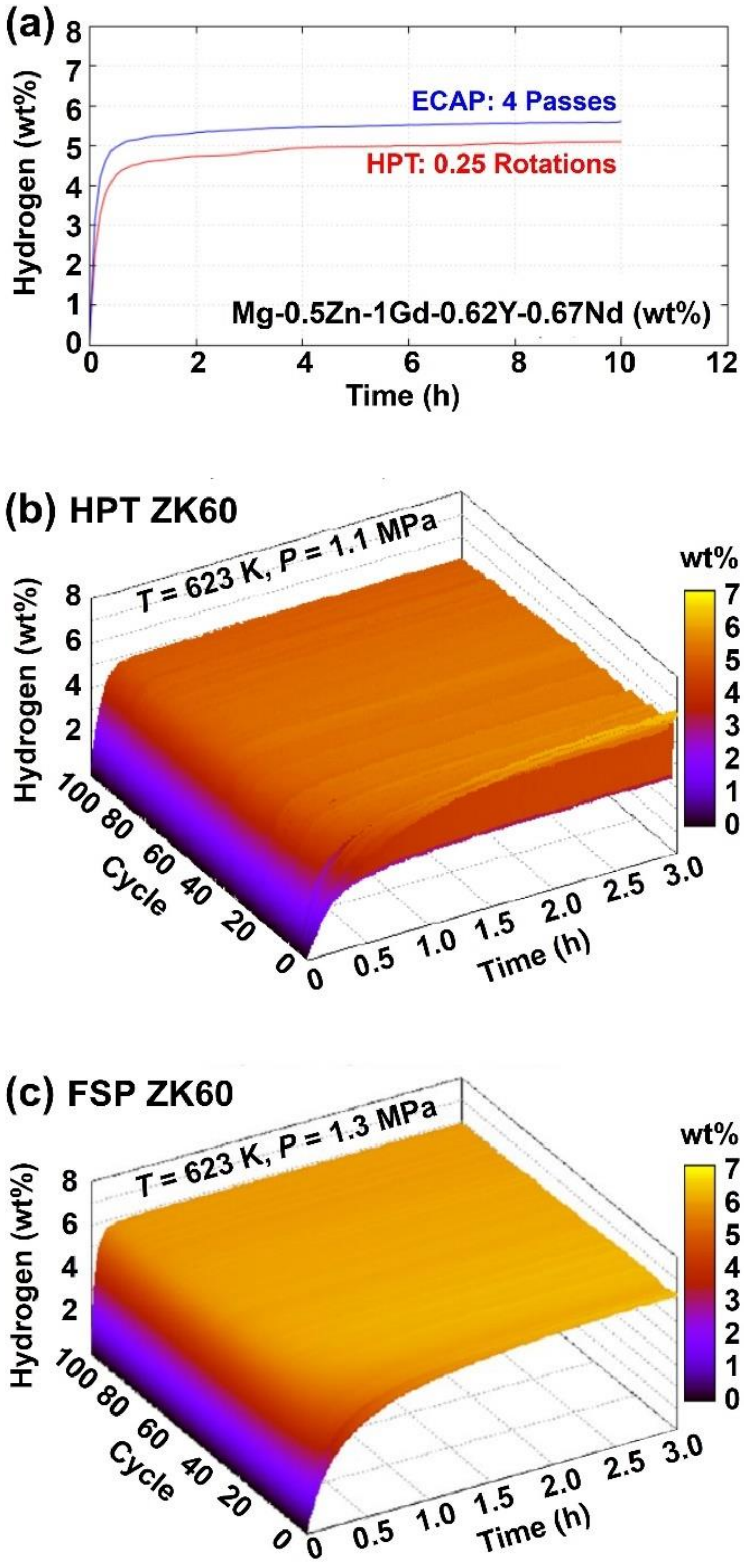

Figure 7. (a) Hydrogen absorption at $T$ = 613 K after ten hydrogenation cycles in Mg-0.5Zn-1Gd-0.62Y-0.67Nd (wt%) alloy processed by HPT and ECAP to von Mises equivalent strain of 4.3 [162]. Cyclic hydrogen absorption kinetic curves at $T$ = 623 K for Mg-based ZK60 alloy processed by (b) HPT and (c) FSP to comparable von Mises equivalent strains of ~7 [161].

### 3.4. Hydrogenation of Inert Materials

We reiterate that SPD is a useful group of processes to produce nano-sized grains decorated with crystal lattice defects [20]. As reviewed in [48,163,164], among the SPD processes, HPT is a unique method that is applicable to hard and brittle materials including intermetallics [165,166], glasses [167,168], ceramics [169,170] and semiconductors [171-174]. In particular, by taking this advantage, HPT was applied to an intermetallic $Mg_2Ni$ compound which gave rise to a significant improvement of the hydrogenation kinetics of the material due to the grain refinement down to the nanoscale [59]. The enhanced kinetics was observed even after annealing of the HPT-processed $Mg_2Ni$, which was associated with the formation of many stacking fault defects [59]. In addition to these kinetic effects, which were discussed earlier in this article, defect generation by HPT processing can lead to hydrogen storage in materials that are otherwise inert to hydrogen. This positive effect in some materials such as TiFe [175], TiV [176] and Ti-Fe-Mn [177] is due to the activation problem having been resolved by HPT processing. In addition, it was shown that the formation of defects by HPT processing can influence the local metal-hydrogen interaction and the hydride formation thermodynamics [178]. It is this thermodynamic advantage that enables the partial absorption of hydrogen in some materials that are normally inert to hydrogen. For example, HPT was successfully applied to store hydrogen in $MgNi_2$ which is considered to be inert to hydrogenation [61]. Some experimental pieces of evidence of the hydrogen absorption capability of $MgNi_2$ are presented below.

Fig. 8a displays X-ray diffraction profiles after HPT processing for $N$ = 2, 5 and 10 turns, including a profile before HPT (labelled by $N$ = 0) [61]. Peak broadening occurred as the number of HPT revolutions increased, indicating that strain was introduced in the sample. The results of the quantitative analysis for hydrogen uptake before and after hydrogenation are plotted in Fig. 8b with respect to the number of HPT turns. The hydrogen content increased with the number of HPT revolutions but remained unchanged for the samples that were not subjected to hydrogenation irrespective of the number of HPT turns. This comparison shows that hydrogen was stored in $MgNi_2$ after the HPT process and the amount increased with the number of HPT turns. The data for crystallite size and anisotropic strain derived from the X-ray diffraction profiles indicated that hydrogen likely stay at lattice defects such as dislocations and grain boundaries but not at the tetragonal sites [61]. Despite the clear evidence that hydrogen could be stored in $MgNi_2$, the amount absorbed was as small as 0.1 wt%. The data presented in Figs. 8a and 8b were obtained six months after the hydrogenation of HPT-processed samples. These low values might be associated with the hydrogen trapped in dislocations or grain boundaries while most of the hydrogen in the crystal lattice was desorbed over the six-month storage. In another study, the amount of absorbed hydrogen was measured using thermo-gravimetric-differential analysis one day after the hydrogenation of HPT-processed samples [179]. The measurements returned a high value of ~1.5 wt% for the weight loss, as shown in Fig. 8c [179]. No change in the X-ray profiles occurred six months after HPT (Fig. 8d), suggesting that a low storage capacity reported in [61] might stem from the desorption of hydrogen while keeping the material for six months in the air. The entirety of the data presented leads one to conclude that HPT has a great potential to activate even inert Mg-based materials for hydrogen storage through modifications of both the thermodynamics and the kinetics of hydrogen sorption.

.

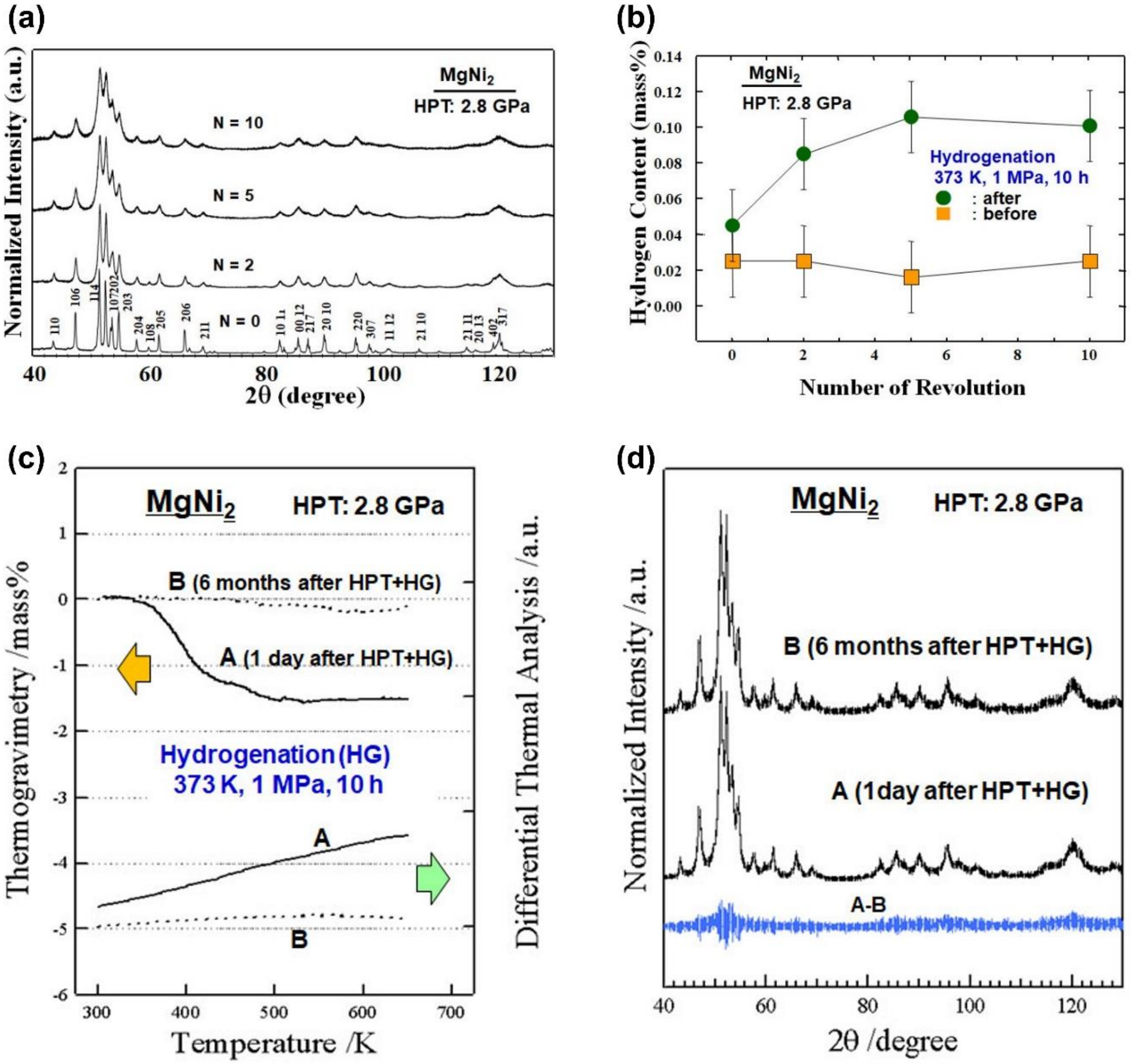

Figure 8. (a) X-ray diffraction profiles after HPT processing for $N$ = 2, 5 and 10 revolutions including profile measured before HPT (labelled $N$ = 0) [61]. (b) Plots of hydrogen content against the number of HPT revolutions, where hydrogenation was made at 373 K under 1 MPa for 24 hours after HPT processing [61]. (c) Thermo-gravimetric differential thermal analysis one day (A) and six months (B) after HPT processing for ten revolutions and hydrogenation [179]. (d) X-ray diffraction profiles measured one day (A) and six months (B) after HPT processing for ten revolutions and hydrogenation including the intensity difference between profiles of A and B (labeled A-B) [179].

## 4. Synthesis of Mg-based Materials with Desirable Thermodynamics

Despite significant progress in the improvement of the kinetics of hydrogenation in magnesium and its alloys by using catalysts or by application of SPD, there has been rather limited success in the development of Mg-based alloys with appropriate thermodynamics for dehydrogenation at room temperature. The main reason for such limited advancements is the large binding energy between the magnesium and hydrogen atoms. Compositional modification is generally considered the most effective strategy to modify the thermodynamics of hydrogen storage in Mg-based alloys [180-183]. The SPD processing does not influence the overall thermodynamics of hydrogen storage in magnesium unless a phase transformation occurs, mechanical alloying is achieved, or small crystals with sizes of a few nanometers are formed. However, the SPD methods and particularly HPT have been used in recent years to synthesize new Mg-based alloys which can store hydrogen at low temperatures or even at room temperature. Some major studies are discussed in this section which include the concept of binding-energy engineering and the developments of nanoglasses, high-entropy alloys and metastable phases.

### 4.1. Binding-Energy Engineering for Room-Temperature Hydrogen Storage

Magnesium hydride $MgH_2$ described above as a promising hydrogen storage material suffers from high dehydrogenation enthalpy (75 kJ/mol $H_2$) due to the strong binding between the magnesium and hydrogen atoms [38,184]. Several major research activities to overcome the key issue of the excessively high thermodynamic stability of $MgH_2$ have been developed: (i) nanoengineering, such as the production of thin films [185-188] and nanoparticles [189-192] including confined [193], non-confined [194], core-shell [195] or composite [196] states; (ii) fabrication of metastable hydrides such as γ-$MgH_2$ [51,98]; and (iii) alloying to form intermetallic compounds, notably $Mg_2Ni$ [197], or solid solutions including Mg-In alloys [198,199]. Among them, alloying is the most feasible way to develop new materials by tuning the chemical composition. However, the thermodynamic immiscibility of Mg in many systems limits the fabrication of Mg-based alloys comprised of three or more component elements, especially by the melting process.

HPT was recently found to be a powerful tool to fabricate new metastable Mg-based alloys with superior hydrogen storage properties at room temperature [19,129]. A successful example is $Mg_4NiPd$, which is expected to have moderate hydrogen binding energy within the $Mg_2Ni$ and $Mg_2Pd$ ones, as predicted by the first-principles calculations (Fig. 9a) [129]. The nominal $Mg_4NiPd$ fabricated by the melting process contained three phases, $Mg_8Ni_3Pd$, $MgNi_2$ and $Mg_5Pd_2$, with heterogeneous elemental distributions (Fig. 9b). The homogeneity of the elemental distribution significantly improved to the atomic scale by 1,500 turns of HPT (Fig. 9b), and as a result, a new phase of $Mg_4NiPd$ with a partly ordered CsCl-type structure was obtained (Fig. 9c). More importantly, at room temperature, the HPT-processed $Mg_4NiPd$ could reversibly absorb and desorb 0.7 wt% of hydrogen (Fig. 9d), which was suggested to occupy the 4Mg-1Ni-1Pd site with a hydrogen binding energy of -0.12 eV (Fig. 9e). These results demonstrated the success of destabilizing the dehydrogenation thermodynamics of Mg-based alloys to room temperature by applying binding energy tailoring principles which could be realized by SPD processing. This strategy can also be applied more universally to design new materials with superior functional properties beyond the scope of the phase equilibria.

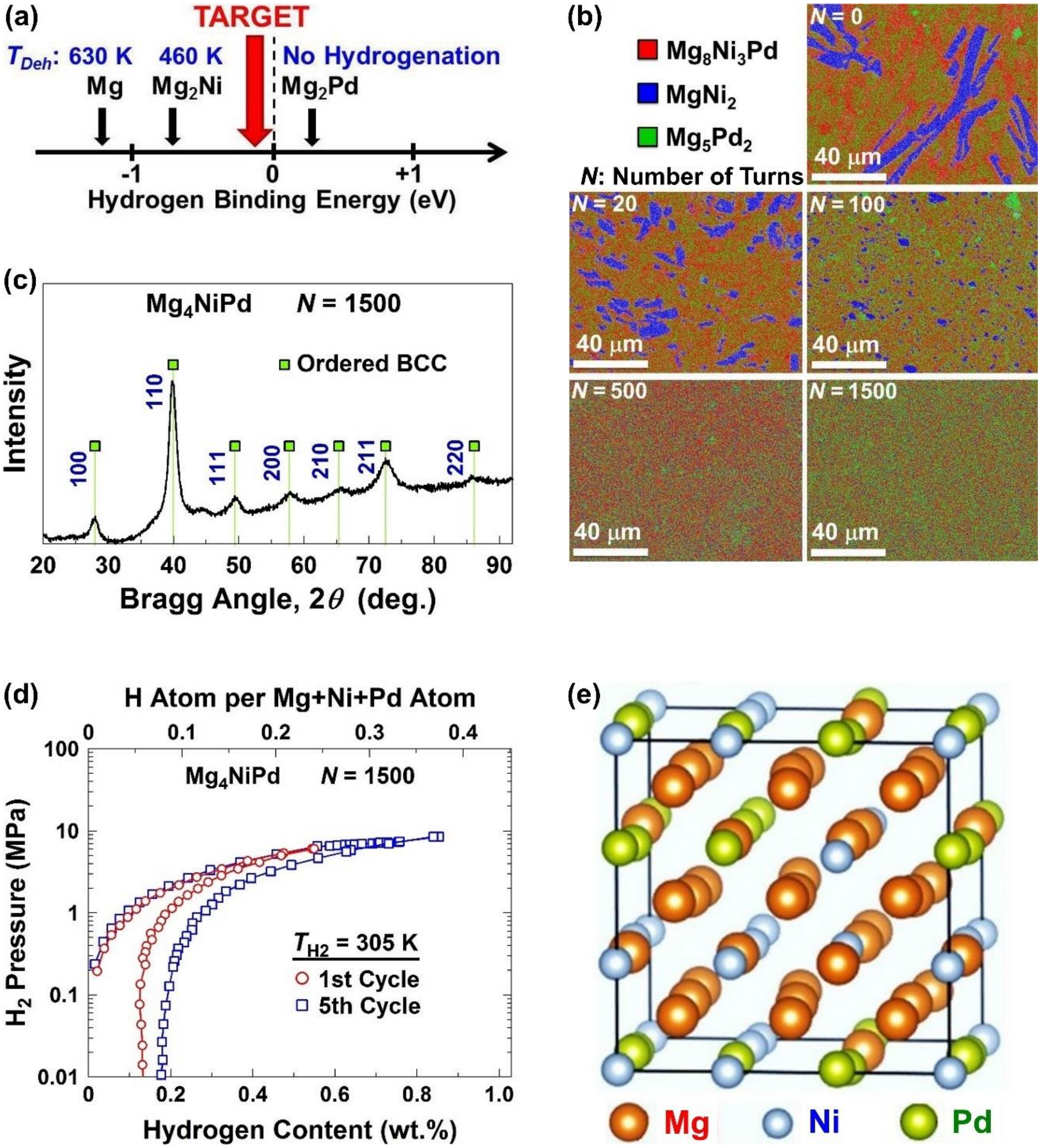

Figure 9. (a) Concept of binding-energy engineering used to design $Mg_4NiPd$. (b) Elemental mapping of $Mg_4NiPd$ after casting ($N$ = 0) and after HPT processing for $N$ = 20, 100, 500 and 1,500 turns. (c) X-ray diffraction profile and (d) pressure-composition isotherm of $Mg_4NiPd$ after 1,500 HPT turns. (e) Crystal structure of homogeneous $Mg_4NiPd$ simulated using first-principles calculations [129].

### 4.2. Nanoglasses

Owing to the thermodynamic stability of $MgH_2$, its hydrogen storage temperature is higher than 573 K [9,184]. Although, as discussed above, nanostructuring is an efficient strategy to improve the hydrogen storage kinetics of $MgH_2$, the thermodynamic stability of nanostructured or nanoconfined $MgH_2$ can hardly be reduced [200-202]. Hydrogen storage properties of Mg-based materials are expected to be varied significantly in their amorphous form compared with their crystalline counterparts. As distinct from the defined chemical compositions in crystalline alloys, amorphous alloys usually have a wider range of chemical compositions which is related to the glass-forming ability of the amorphous alloys [203]. Therefore, the tunability of hydrogen storage properties in amorphous alloys is supposed to be much wider than that in crystalline alloys.

In the past two decades, various Mg-based amorphous alloys have been prepared by melt spinning or ball milling and studied as hydrogen storage materials. These includes Mg-Ni [204], Mg-Ni-Y [205], Mg-Ni-RE [206], Mg-Ni-Fe [207], Mg-Ni-La [208], Mg-Mm-Ni [209,210], Mg-Y-Ni [211], $Mg_{65}Ni_{20}Cu_5Y_{10}$ [70], $Mg_{90}Ni_8RE_2$ (RE = Y, Nd, Gd) [212], $Mg_{85}Ni_{15-x}M_x$ (M = Y or La, $x$ = 0 or 5) [213], $Mg_{100-x}Ni_x$ alloys ($x$ = 0.5, 1, 2, 5) [214], $LaMg_{11}Ni$ [215], Mg-Ce-Ni [216], $RMg_2Ni$ (R = La, Ce, Pr, Nd) [217], $Mg_{11}Y_2Ni_2$ [218] and $Mg_{80}Ce_{10}Ni_{10}$ [219]. However, most studies showed that the amorphous alloys would decompose into crystalline alloys or hydrides upon hydrogenation/dehydrogenation [204-219], and thus the hydrogen storage reversibility of amorphous alloys is usually not satisfactory. It was reported that the Mg-Ce-Ni amorphous alloys could reversibly absorb and desorb 0.3 wt.% of hydrogen at room temperature [220]; however, the reversible hydrogen storage of this alloy could not be improved by increasing the temperature up to 423 K [221]. Therefore, improving the reversible hydrogen storage capacity of Mg-based amorphous alloys at low temperatures remains a challenging issue.

SPD was shown to be beneficial for improving the hydrogen storage properties of Mg-based amorphous materials, particularly for enhancing hydrogen storage kinetics and reducing hydrogen storage temperature. For example, high-pressure calorimetry measurements revealed that hydrogen uptake in the fully amorphous $Mg_{65}Ni_{20}Cu_5Y_{10}$ alloy, prepared by melt spinning followed by HPT processing, occurs at a significantly lower temperature compared to the fully crystallized state [70]. In another trial, the HPT technique was used to process the melt-spun $Mg_{65}Ce1_0Ni_{20}Cu_5$ amorphous alloy to produce a nanoglass alloy [69]. Fig. 10a shows that the X-ray diffraction pattern of the melt-spun $Mg_{65}Ce1_0Ni_{20}Cu_5$ alloy is typically amorphous, while the HPT-treated samples contain a small fraction of crystalline phases. Fig. 10b demonstrates that superior hydrogen storage could be achieved in these HPT-treated alloys at a low temperature of 393 K, while the melt-spun sample was not active at this temperature. It was suggested that the HPT-induced nanoglass formation (i.e. amorphous phase with a large fraction of interfaces at the nanometer scale) leads to a large number of hydrogen diffusion channels and hydrogen storage sites for the improvement of hydrogenation properties (Fig. 10c). Such hydrogen storage performance in the HPT-processed sample is comparable with the performance of thin films of $Mg_{65}Ce1_0Ni_{20}Cu_5$. As shown in Figs. 10d and 10e, the amorphous thin films with thicknesses of 50-300 nm could reversibly absorb and desorb hydrogen with almost unchanged desorption kinetics at a low temperature of 393 K [222]. Therefore, SPD-processed nanoglasses and nanosized amorphous thin films are promising candidates for hydrogen storage at moderate temperatures, but nanoglasses can be produced by SPD in the bulk form with a larger size which is beneficial for practical applications.

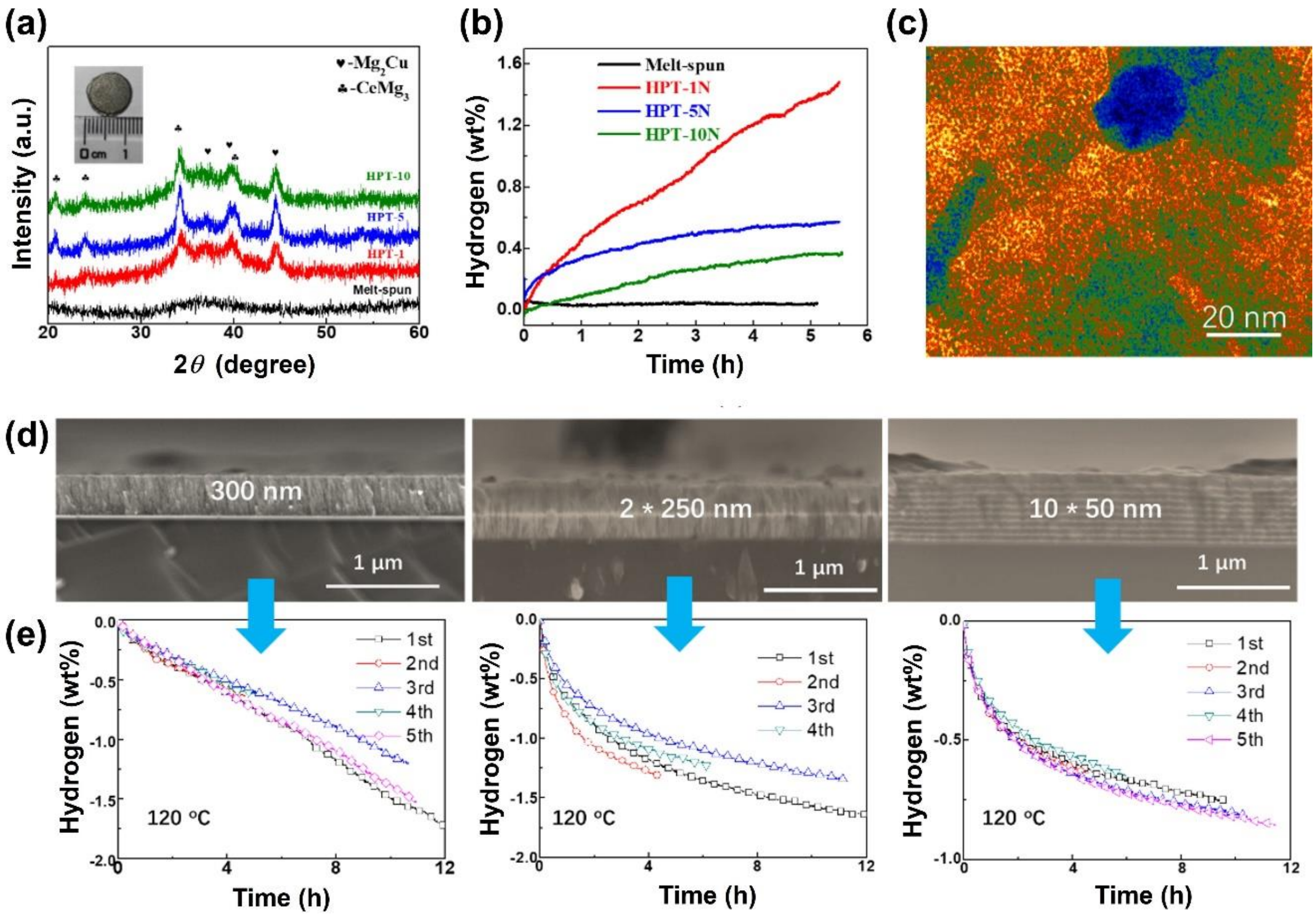

Figure 10. Hydrogen storage in $Mg_{65}Ce_{10}Ni_{20}Cu_5$ nanoglass and amorphous thin films. (a) X-ray diffraction patterns of the melt-spun alloy before and after HPT processing for 1, 5 and 10 turns. Inset: an image of a disc-shaped specimen after the HPT process [69]. (b) Hydrogenation kinetics of melt-spun and HPT-treated samples [69]. (c) High-resolution image of sample processed by one HPT turn taken by transmission electron microscopy [69]. (d) Cross-sectional images of films with thicknesses of 300, 250 and 50 nm taken by scanning electron microscopy [222]. (e) Hydrogen desorption kinetics of amorphous thin films [222].

### 4.3. High-Entropy Alloys

In a quest for new materials for hydrogen storage, multicomponent and high-entropy alloys and their corresponding hydrides have attracted wide attention due to the huge compositional field that can be accessed and the unexpected potentialities of hydrogen storage properties they promise [223,224]. The vast compositional field offers virtually endless opportunities for designing and/or discovering new alloys with superior hydrogen storage properties [225]. The concept of high-entropy materials was first proposed in 2004 by independent studies by Yeh *et al.* [226] and Cantor *et al.* [227]. They introduced a new family of metallic alloys that was based on a mixture of at least five principal elements, each having an atomic percentage in the range of 5 at% to 35 at%. The presence of multiprincipal elements leads to the stabilization of single-phase solid solutions with main crystal structures of body-centered cubic (BCC), face-centered cubic (FCC) or hexagonal close-packed (HCP). High-entropy hydrides also consist of at least five principal elements and hydrogen, as shown in Fig. 11a [228].

Several high-entropy alloys for hydrogen storage have been proposed mainly including refractory elements such as TiVZrNb [229], TiZrNbTa [230], TiVZrNbHf [231,232], TiZrNbHfTa [233,234], TiZrNbFeNi [235] and TiZrNbCrFe [236]. Despite the remarkable results observed in hydrogen storage properties so far, the gravimetric storage capacities of these alloys are relatively low due to the high atomic weight of refractory elements (zirconium, hafnium, niobium, tantalum). Accordingly, one of the strategies to solve this limitation is to proceed with the substitution of refractory elements with lightweight elements with a high affinity to hydrogen such as magnesium. To that end, Mg-containing high-entropy alloys have been explored for hydrogen storage such as $MgZrTiFe_{0.5}Co_{0.5}Ni_{0.5}$ [237], $MgTiNbCr_{0.5}Mn_{0.5}Ni_{0.5}$ [238], MgVAlCrNi [239], $Mg_{12}Al_{11}Ti_{33}Mn_{11}Nb_{33}$ [240], $Mg_{35}Al_{15}Ti_{25}V_{10}Zn_{15}$ [241] and MgVTiCrFe [242]. These Mg-containing high-entropy alloys show the presence of single solid-solution phases and are mainly prepared by high-energy ball milling. The SPD techniques such as HPT have also been successfully used to synthesize new Mg-containing multicomponent alloys [243] and high-entropy alloys [68], although the hydrogen storage performance of these alloys is not still comparable with the Ti-Zr-containing high-entropy alloys with the Laves phase structure [228,243].

Among various SPD methods, HPT is currently the most popular one to synthesize multicomponent and high-entropy alloys [244]. MgVCr is a Mg-based multicomponent hydrogen storage alloy with the BCC structure that was synthesized by HPT [242]. As shown in Fig. 11b, the application of HPT to a powder mixture of magnesium, vanadium and chromium, could lead to the formation of a new alloy with a single BCC phase. This multicomponent alloy stored up to 0.9 wt% of hydrogen at ambient temperature, but the reversibility of the storage was rather poor. The MgTiVCrFe alloy is a Mg-containing high-entropy alloy that was synthesized by ball milling followed by HPT processing [68]. The alloy showed a BCC solid solution phase together with a high level of amorphization. The hydrogen uptake capacity of alloy MgTiVCrFe at 623 K was found to be 0.3 wt%, as seen in Fig. 11c. The outcomes of the investigations conducted so far indicate that SPD processing is a possible pathway to synthesize Mg-containing high-entropy alloy. However, this research needs to be empowered by developing theoretical concepts for material design, as attempted earlier for Ti-containing high-entropy hydrogen storage materials [228,243].

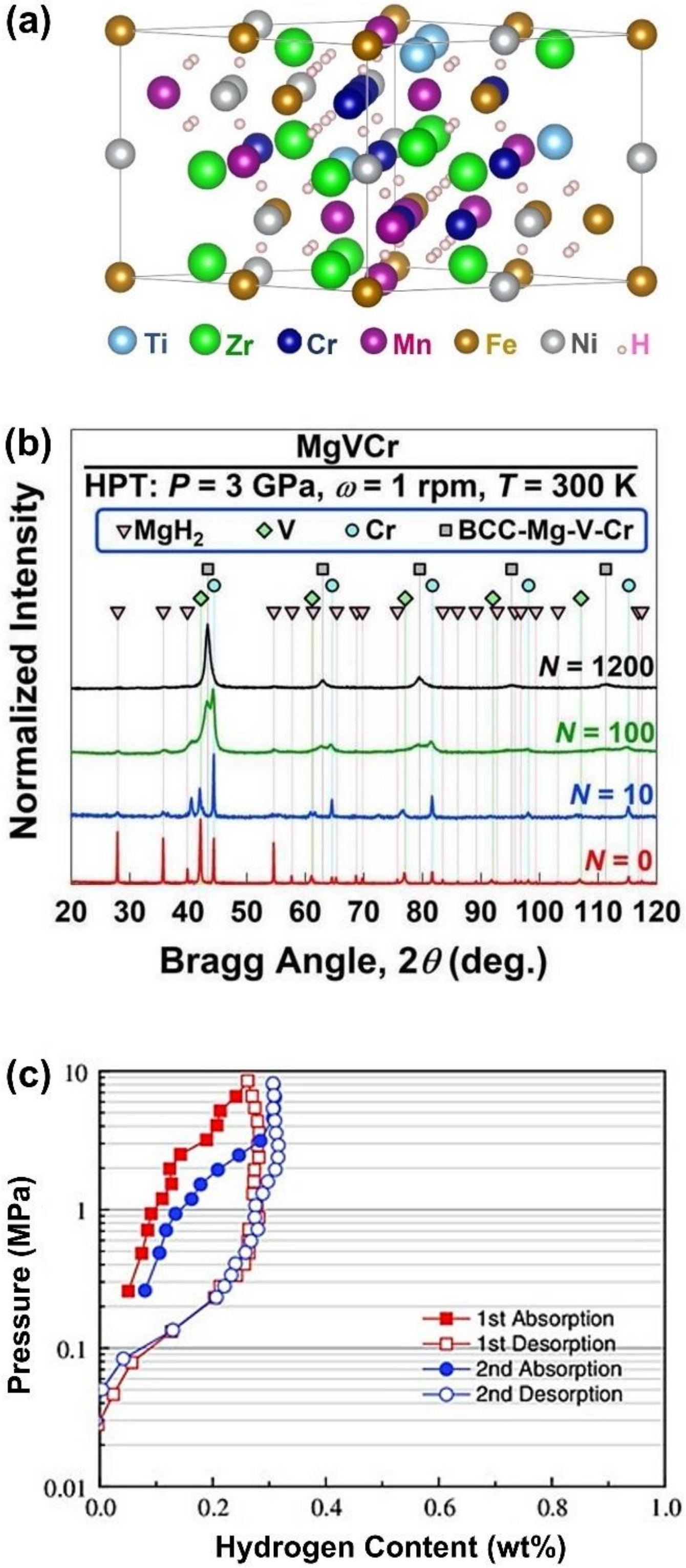

Figure 11. (a) Typical structure of high-entropy hydrides [243]. (b) X-ray diffraction profiles of a mixture of magnesium, vanadium and chromium powders before and after HPT processing for $N$ = 10, 100 and 1,200 revolutions [242]. (c) Hydrogen pressure-composition isotherms at 623 K for MgTiVCrFe synthesized by ball milling followed by HPT processing [68].

### 4.4. Metastable Phases

Following a publication in 1988 on the application of HPT to an aluminum alloy, the main goal of SPD processing has become the production of ultrafine-grained materials [245]. However, a survey in the classic publications indicates that the first SPD methods were invented to control polymorphic phase transformations under high pressure and large shear strain [246]. Nowadays, the SPD methods and particularly HPT are still used to control polymorphic phase transformations [247-249] and solid-state reactions [250,251]. The main reason for a significant interest in using HPT to control phase transformation is the high applicable pressure in this method which can lead to the synthesis of high-pressure phases [247,249]. Moreover, the shear strain in this method can be increased almost without any limitations, a fact that was used to employ ultra-SPD for synthesizing new materials by mechanical alloying through HPT [46]. Moreover, a combination of high pressure and high strain in this method permits for synthesizing hard and brittle materials such as TiFe hydrogen storage material [252]. This capability of HPT was successfully used to synthesize various Mg-X alloys and $Mg_2X$ intermetallics (X: 21 different elements) [253], while the synthesis of Mg-based compounds by melting techniques is usually hard. It is remarkable that HPT was recently used to synthesize new metastable hydrogen storage materials even from immiscible systems such as Mg-Ti [63,98], Mg-Zr [64], Mg-Hf [254], Mg-V-Cr [242] and Mg-Ni-Pd [129]. As discussed earlier, some of these metastable alloys such as $Mg_4NiPd$ exhibit thermodynamics suitable for reversible room-temperature hydrogen storage, while some others such as MgHf show poor reactivity with hydrogen [254]. Here, the impact of HPT on the production of the high-pressure $\gamma$-$MgH_2$ phase with low thermodynamic stability is discussed [51].

As shown in Fig. 12a, $MgH_2$ has an $\alpha$ phase with the tetragonal structure at ambient pressure and shows transitions to a $\gamma$ phase with the orthorhombic structure at 0.39-5.5 GPa and to a $\beta$ phase with the cubic structure at 3.9-9.7 GPa [255]. Some studies suggested that the partial formation of $\gamma$ phase after mechanical milling [256], electrochemical process [257] and plasma sputtering [258] can result in the destabilization of hydride for easy hydrogen desorption. To examine this issue, HPT was recently used to synthesize almost 100% of $\gamma$-$MgH_2$ and examine its dehydrogenation behavior [51]. As shown in the X-ray diffraction profiles of Fig. 12b, a transition from the $\alpha$ phase to the $\gamma$ phase occurs by HPT processing under 5 GPa, while the fraction of the high-pressure phase increases with increasing the number of HPT turns (i.e. increasing strain). Examination of lattice images by high-resolution transmission electron microscopy, shown in Figs. 12c and 12d, also indicates the formation of large amounts of $\gamma$ phase (almost 100%) after HPT processing for $N$ = 15 turns. Differential scanning calorimetry and thermogravimetry analyses, shown in Figs. 12e and 12f, clearly confirm a reduction in the dehydrogenation temperature by increasing the fraction of the $\gamma$ phase. These results confirm that the production of metastable $\gamma$-$MgH_2$ is a strategy to reduce thermodynamic stability, as suggested by first-principles calculations in [51]. However, the reversibility of this phase is another issue that should be addressed by adjusting the chemical composition, as attempted for some other metastable phases such as $Mg_4NiPd$ [129].

In conclusion, HPT appears as an effective SPD method to synthesize metastable alloys and hydrides for hydrogen storage. However, the method has two drawbacks for practical applications. First, the sample in HPT is in the form of a disc with small sizes. Second, strain is not uniform along the disc radius resulting in the nonuniform distribution of metastable phases. There are now some signs of progress in upscaling the sample size in HPT [163], which can open a path for future applications of the method. Moreover, it was suggested that using ring samples rather than disc samples can diminish the heterogeneity in the final product [259].

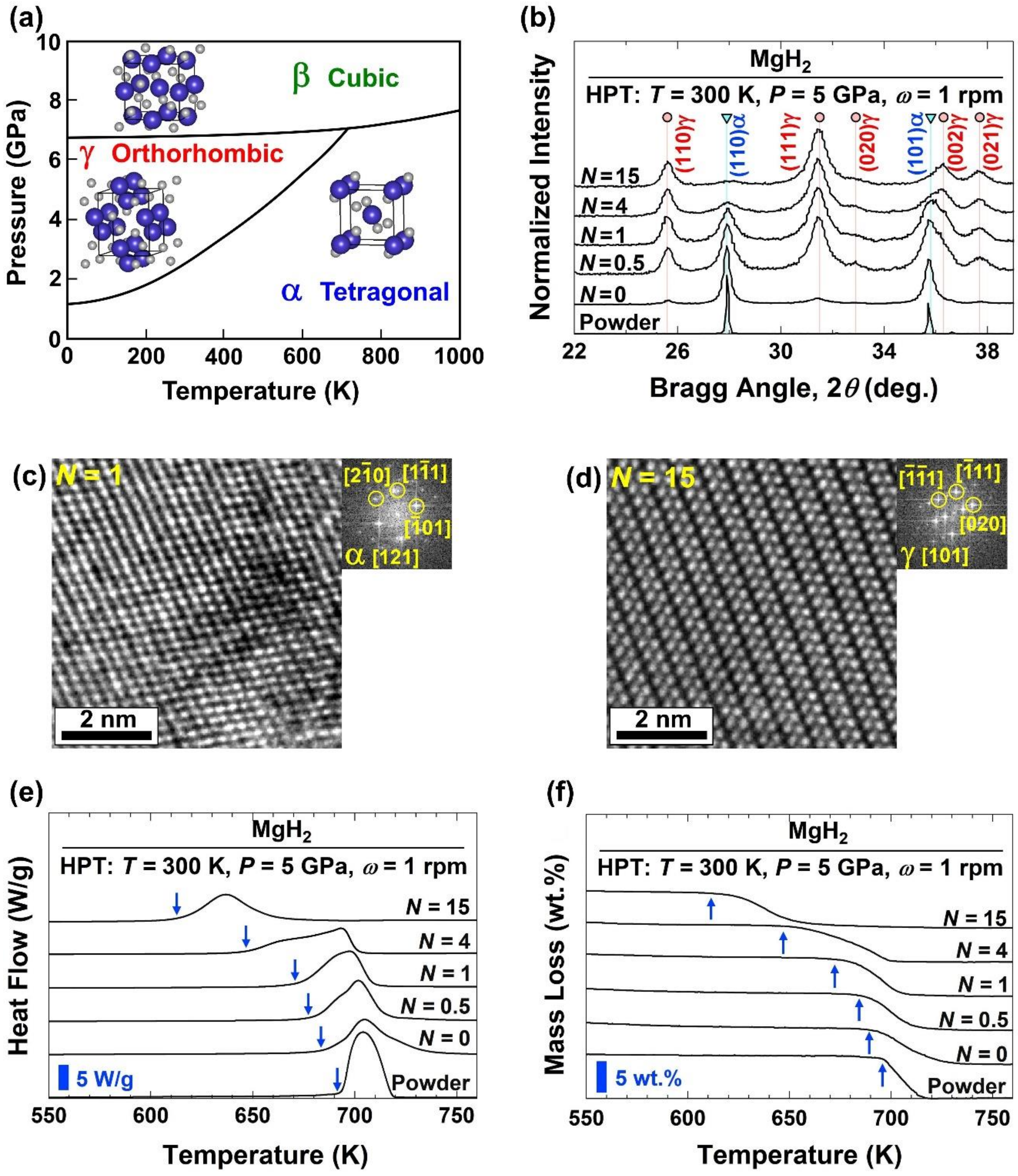

Figure 12. (a) Pressure-temperature phase diagram of magnesium hydride. (b) X-ray diffraction profiles of $MgH_2$ processed by HPT for various turns ($N$). High-resolution lattice images of $MgH_2$ after HPT processing for (c) 1 and (d) 15 turns. (e) Heat flow in differential scanning calorimetry and (f) mass loss in thermogravimetry for $MgH_2$ processed by HPT for various turns [51].

## 5. Concluding Remarks and Outlook

Severe plastic deformation methods are receiving significant attention in the context of the processing of hydrogen storage materials. The initial motivation to apply these processes to magnesium and its alloys was the improvement of the kinetics of hydrogen storage, but currently, there are numerous reports on the application of these methods to produce novel materials with appropriate thermodynamics and low-temperature hydrogen storage capability. Currently, a wide range of processes including equal-channel angular pressing, high-pressure torsion, intensive rolling and fast forging are employed for processing Mg-based hydrogen storage materials, as discussed in Section 2. While high-pressure torsion is the most powerful technique for fundamental studies, the three other methods have a greater potential for commercial applications. The application of severe plastic deformation introduces ultrafine grains and high densities of crystal lattice defects in bulk samples, and this results in better activation, high air resistance, fast hydrogen absorption/desorption kinetics, and appropriate cycling stability, as discussed in Section 3. It was shown that some materials which are apparently inert to hydrogen become active after processing and can store hydrogen. The large shear strain and high pressure involved in severe plastic deformation enable the synthesis of new materials such as nanoglasses, high-entropy alloys and metastable phases with thermodynamics amenable for hydrogen storage, as discussed in Section 4. A combination of theoretical binding-energy engineering and severe plastic deformation can successfully lead to achieving reversible hydrogen storage in Mg-based alloys even at room temperature.

The findings reviewed in this article introduce severe plastic deformation as a strong tool to produce advanced hydrogen storage materials. However, there are several issues that need to be addressed in the future. (i) Scaling up the processes is still a key issue for commercialization. (ii) Heat transfer is of paramount significance in hydrogen storage materials, but so far this aspect has not been sufficiently examined for severely deformed magnesium. (iii) Cycling stability of severely deformed Mg-based hydrogen storage materials requires further investigation. (iv) Addition of catalysts in conjunction with SPD processing may be a potential pathway to enhancing the hydrogenation/dehydrogenation kinetics at low temperatures. (v) More importantly, further studies on the combination of thermodynamic calculations and experiments are needed to discover new materials that can satisfy the desired requirements for stationary and mobile hydrogen storage applications. By considering the worldwide trend to realize carbon-neutral energy and utilize hydrogen as a zero $CO_2$ emission fuel, it is expected that in the future, severe plastic deformation technologies will contribute to the hydrogen economy more prominently.

**Acknowledgments**

K.E. was supported in part by the Light Metals Educational Foundation of Japan, and in part by the MEXT, Japan through Grants-in-Aid for Scientific Research on Innovative Areas (JP19H05176 & JP21H00150) and Challenging Research Exploratory (JP22K18737).

W.J.B. is grateful to the Brazilian agencies FAPESP (grant number 2013/05987-8) and CNPq (grant number 421181-2018-4 and 307397-2019-0) for the financial support and to the Laboratory of Structural Characterization (LCE-DEMa-UFSCar) for general electron microscopy facilities.

R.F. thanks for the financial support from FAPESP (grant number 2022/01351-0).

T.G. thanks the support from the French State through the ANR-21-CE08-0034-01 project as well as the program "Investment in the future" operated by the National Research Agency (ANR) and referenced under No. ANR-11-LABX-0008-01 (Labex DAMAS).

H.L. appreciates the financial support from the National Natural Science Foundation of China (grant number 52171205).

H.J.L. thanks the financial support from the National Natural Science Foundation of China (grant number 52071157).